\documentclass[12pt,a4paper]{article}

\usepackage[a4paper,left=2.0cm,right=2.0cm, bottom=2.5cm]{geometry}
\usepackage{ifthen}
\usepackage{subfig}
\usepackage{graphicx}
\usepackage{longtable}
\usepackage[abs]{overpic}
\usepackage{multirow}
\usepackage{braket}
\usepackage{lipsum}
\usepackage{comment}
\usepackage{amsmath}
\usepackage{amssymb}
\usepackage{siunitx}
\usepackage{booktabs}
\usepackage{lineno}

\usepackage{color}   
\usepackage{hyperref}
\hypersetup{linktocpage}
\hypersetup{
    colorlinks=true, 
    citecolor=blue,
    linkcolor=blue,     
}

\newboolean{pdflatex}

\newboolean{uprightparticles}
\setboolean{uprightparticles}{false} 

\newcommand*{\GeVc} {\ensuremath{\text{Ge\kern -0.1em V/}c} }

\cleardoublepage

\begin{document}

\begin{titlepage}

\vspace*{-3.5cm}

\hspace*{-0.5cm}
\begin{tabular*}{\linewidth}{lc@{\extracolsep{\fill}}r}

\vspace*{-12mm}\mbox{\includegraphics[width=0.15\textwidth]{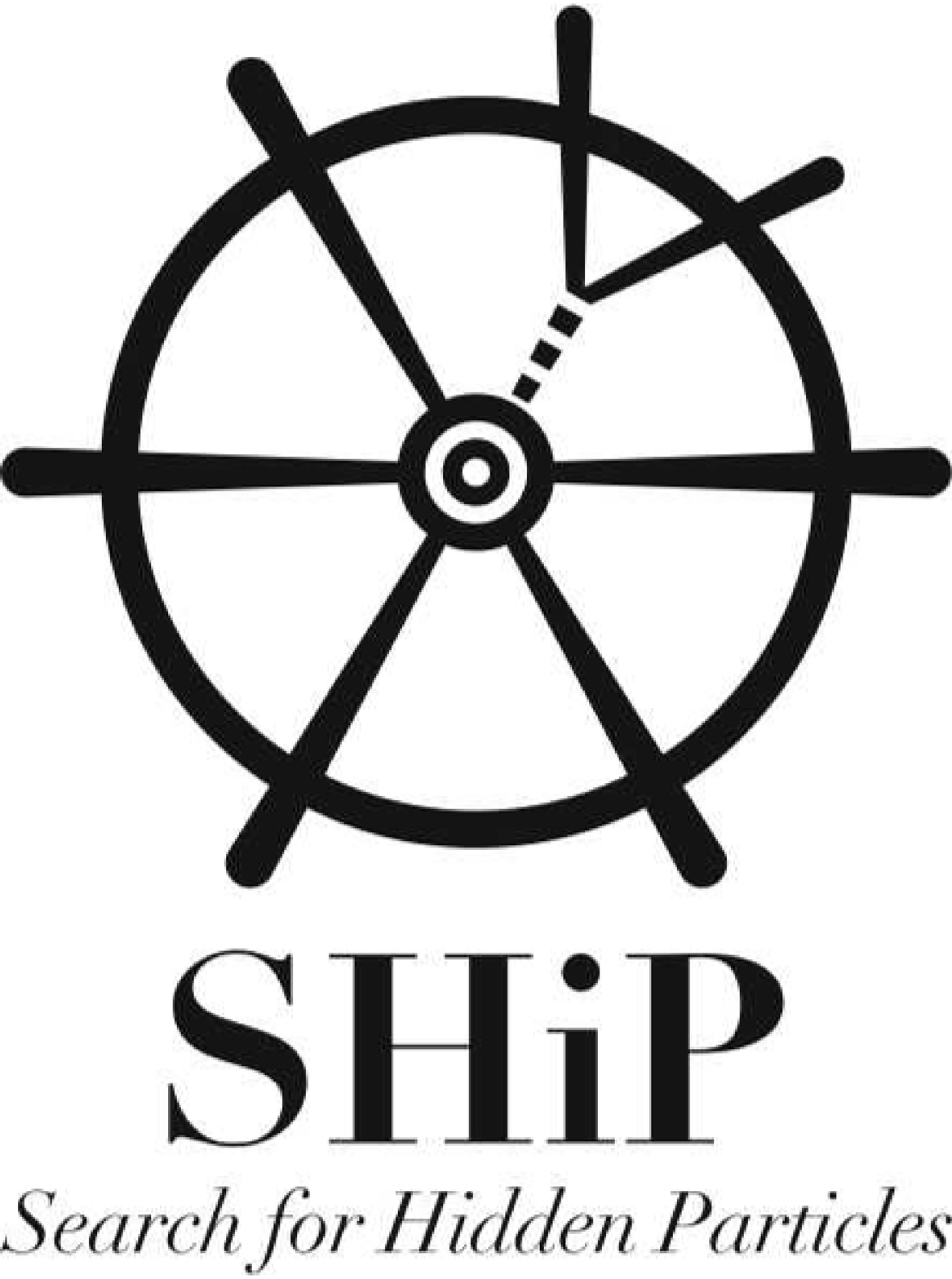}} & \\
 & & \today \\ 
 & & \\
\hline
\end{tabular*}

\vspace*{4.0cm}

{\bf\boldmath\huge
\begin{center}
{\bf\huge\boldmath SND@LHC}
\end{center}
}

\vspace*{2.0cm}

\begin{center}
SHiP Collaboration
\bigskip\\
\end{center}

\begin{abstract}
We propose to build and operate a detector that, for the first time, will measure the process $p p \rightarrow \nu X$ at the LHC and search for feebly interacting particles (FIPs) in an unexplored domain. The TI18 tunnel has been identified as a suitable site to perform these measurements due to  very low  machine-induced background. The detector will be off-axis with respect to the ATLAS interaction point (IP1) and, given the pseudo-rapidity range accessible,  the corresponding neutrinos will mostly come from charm decays: the proposed experiment will thus make the first test of the heavy flavour production in a  pseudo-rapidity range that is not accessible by the current LHC detectors. In order to efficiently reconstruct neutrino interactions and identify their flavour, the detector will combine in the target region nuclear emulsion technology with scintillating fibre tracking layers and it will adopt a muon identification system based on scintillating bars that will also play the role of a hadronic calorimeter. The time of flight measurement will be achieved thanks to a dedicated timing detector. The detector will be a small-scale prototype of the scattering and neutrino detector (SND) of the SHiP experiment: the operation of this detector will provide an important test 
of the neutrino reconstruction in a high occupancy environment.
\end{abstract}
\end{titlepage}

\cleardoublepage



%
\centerline{\Large\bf The SHiP Collaboration}
\vspace*{1mm}
\begin{flushleft}
C.~Ahdida$^{44}$,
A.~Akmete$^{48}$,
R.~Albanese$^{14,d,h}$,
A.~Alexandrov$^{14,32,34,d}$,
M.~Andreini$^{44}$,
A.~Anokhina$^{39}$,
S.~Aoki$^{18}$,
G.~Arduini$^{44}$,
E.~Atkin$^{38}$,
N.~Azorskiy$^{29}$
% ...

\vspace*{1cm}

{\footnotesize \it

% ...
$ ^{14}$Sezione INFN di Napoli, Napoli, Italy\\
% ...
$ ^{18}$Kobe University, Kobe, Japan\\
% ...
$ ^{29}$Joint Institute for Nuclear Research (JINR), Dubna, Russia\\
% ...
$ ^{32}$P.N.~Lebedev Physical Institute (LPI RAS), Moscow, Russia\\
% ...
$ ^{34}$National University of Science and Technology "MISiS", Moscow, Russia\\
% ...
$ ^{38}$National Research Nuclear University (MEPhI), Moscow, Russia\\
$ ^{39}$Skobeltsyn Institute of Nuclear Physics of Moscow State University (SINP MSU), Moscow, Russia\\
% ...
$ ^{44}$European Organization for Nuclear Research (CERN), Geneva, Switzerland\\
% ...
$ ^{48}$Middle East Technical University (METU), Ankara, Turkey\\
% ...
$ ^{d}$Universit\`{a} di Napoli ``Federico II'', Napoli, Italy\\
% ...
$ ^{h}$Consorzio CREATE, Napoli, Italy\\
% ...
}
\end{flushleft}


\clearpage

\section{Physics motivation}
\label{sec:phmotivation}

The Standard Model of particle physics (SM) has been able to describe all known microscopic physics phenomena with great precision. However, as well known, there are several failures of the SM of cosmological and astrophysical origin, such as the existence of Dark Matter (DM) and the baryon-antibaryon asymmetry in the Universe, along with neutrino masses and oscillations that also require some physics beyond the Standard Model. 

To-date no firm signs of new physics that would provide resolution to these puzzles have been discovered at the LHC and other particle physics experiments. 
Recently, measurements from LHCb and B-factories have shown a set of discrepancies with respect to SM predictions (see e.g.~\cite{Lees:2012xj,Aaij:2014ora,Aaij:2015yra}). While it is not clear yet if these measurements are statistical fluctuations or a genuine sign of new physics, the experiments, such as ATLAS, CMS, LHCb and Belle, will continue to test them in a canonical way.
It is therefore important to perform complementary measurements. Below we describe how to test lepton flavor universality in the neutrino sector. 

The neutrino sector  
allows precise tests of the SM~\cite{Brock:1993sz,Conrad:1997ne,Formaggio:2013kya,Lellis:2004yn} and a probe for new physics~\cite{Marfatia:2015hva,Arguelles:2019xgp}, in an otherwise veiled region of the Universe~\cite{Ackermann:2019ows}. 
Neutrino interactions have been measured in the energy regime below 350GeV, and recently, the IceCube collaboration reported a few tens of events in the region 10~TeV-1~PeV~\cite{Aartsen:2017kpd}. It is 
notable that the region between 350~GeV and 10~TeV is currently unexplored. 
Measurements of neutrino interactions in the last decades were mainly performed at energies where neutrino oscillations are enhanced over available baselines. 
In addition, there are basically no electron neutrino measurements above 10GeV. Concerning tau neutrinos, only an handful of events observed by DONUT~\cite{Kodama:2007aa} and OPERA~\cite{Acquafredda:2009zz,Agafonova:2018auq} experiments have been recorded. 
 The DONUT experiment at Fermilab performed the first observation of tau neutrinos using a 800~GeV proton beam dump on to a tungsten target. The experiment observed the first nine tau neutrino candidates~\cite{Kodama:2007aa}, without distinguishing between neutrinos and anti-neutrinos and did not study interactions of other neutrino types. OPERA observed ten tau neutrino candidates coming from muon neutrino oscillations and also detected  35 electron neutrino interactions~\cite{Agafonova:2018dkb}. The emulsion technology developed by OPERA proved to be capable of identifying the three neutrino flavours. 

Recently, a beam dump facility has been designed to host the SHiP experiment~\cite{Anelli:2015pba}, searching for hidden particles and studying neutrino physics~\cite{Alekhin:2015byh}. Beam dump facilities provide an intense flux of neutrinos when using a correspondingly intense proton beam. If the proton has sufficient energy, the neutrino beam includes also tau neutrinos. This is the case for SHiP where, in about a decade from now, $2 \times 10^{20}$  protons from the SPS at CERN with 400\,GeV energy will be used. The corresponding neutrino energies will therefore still be limited to 350\,GeV, although the flux will be unprecedented. A beam dump at the LHC could extend the energy range to the TeV scale, but the technology to operate such a facility is not presently available. 
Our proposal aims at measurement of neutrino properties in this previously unexplored domain above of 100s of GeV.
The same experimental setup also allows to pursue another important scientific goal -- searching for \textit{feebly inteacting particles} with masses in the GeV range.

\bigskip

Feebly Interacting  Particles (FIPs) searches are currently undergoing an explosion of interest~\cite{Strategy:2019vxc}. This interest is also stimulated by the lack of any discovery in various direct detection dark matter experiments, which have reached unprecedented levels of precision,  as well as in collider searches which have collected large data samples with increasingly sophisticated analysis techniques. The majority of these efforts are targeting dark particle masses from about 10 GeV to 1 TeV, leaving the lower mass range much less explored. The scarce sensitivity at lower masses is related to the difficulty of detecting the corresponding very soft recoils, given the expected non-relativistic nature of the galactic dark matter. 

On theoretical side, these developments also stimulated significant activity.
Many models have been proposed which expand the interesting FIPs mass range to lower values (GeV and below), while still giving rise to the expected relic DM abundance.  One of the most prominent concepts is the existence of a ``dark sector'', which communicates with the SM particles through a feebly coupled mediator (see e.g.~\cite{Alekhin:2015byh,Strategy:2019vxc}).

\begin{figure}[h!]
\centering
\includegraphics[height=0.3\linewidth]{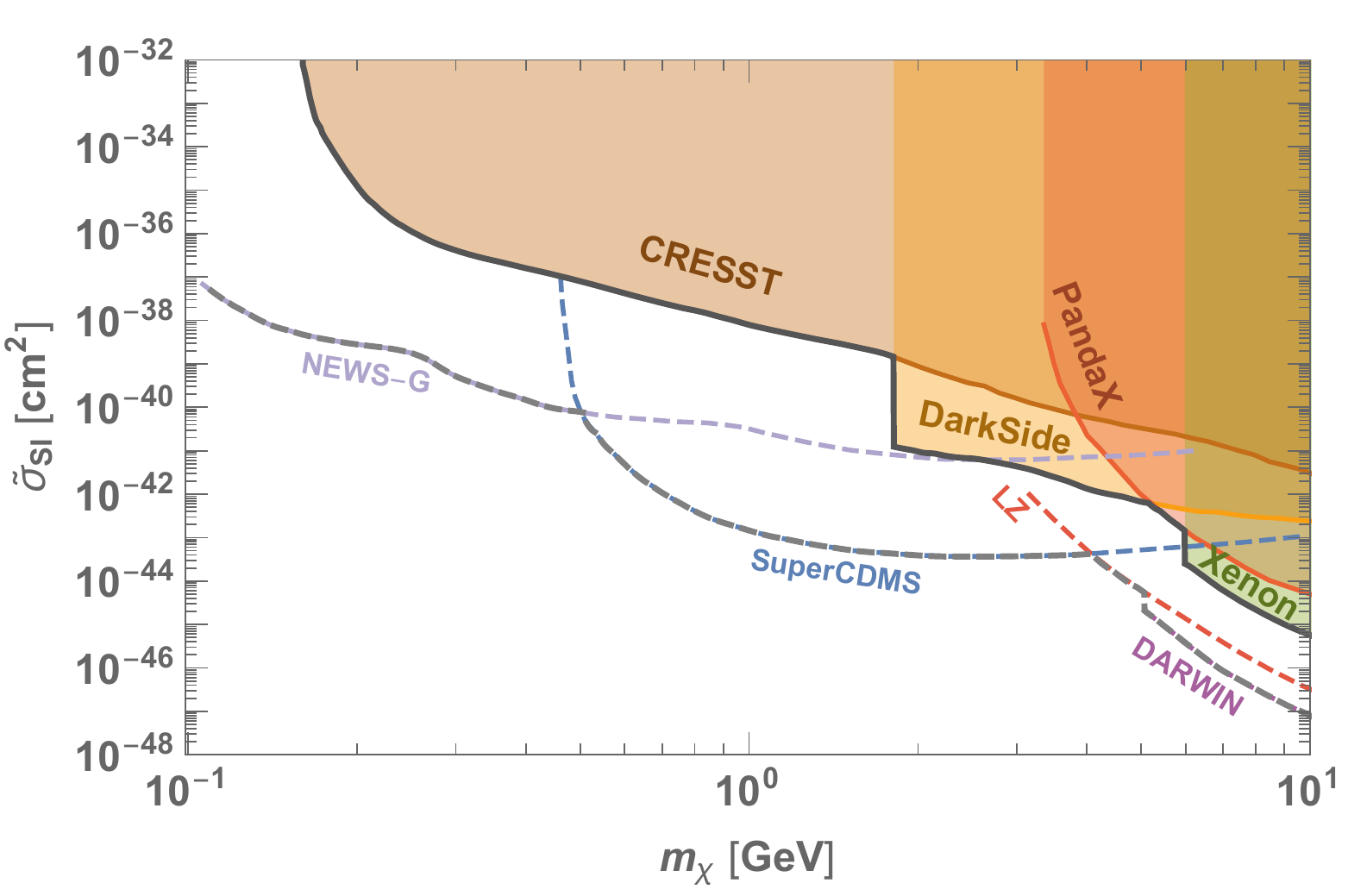}
\includegraphics[height=0.3\linewidth]{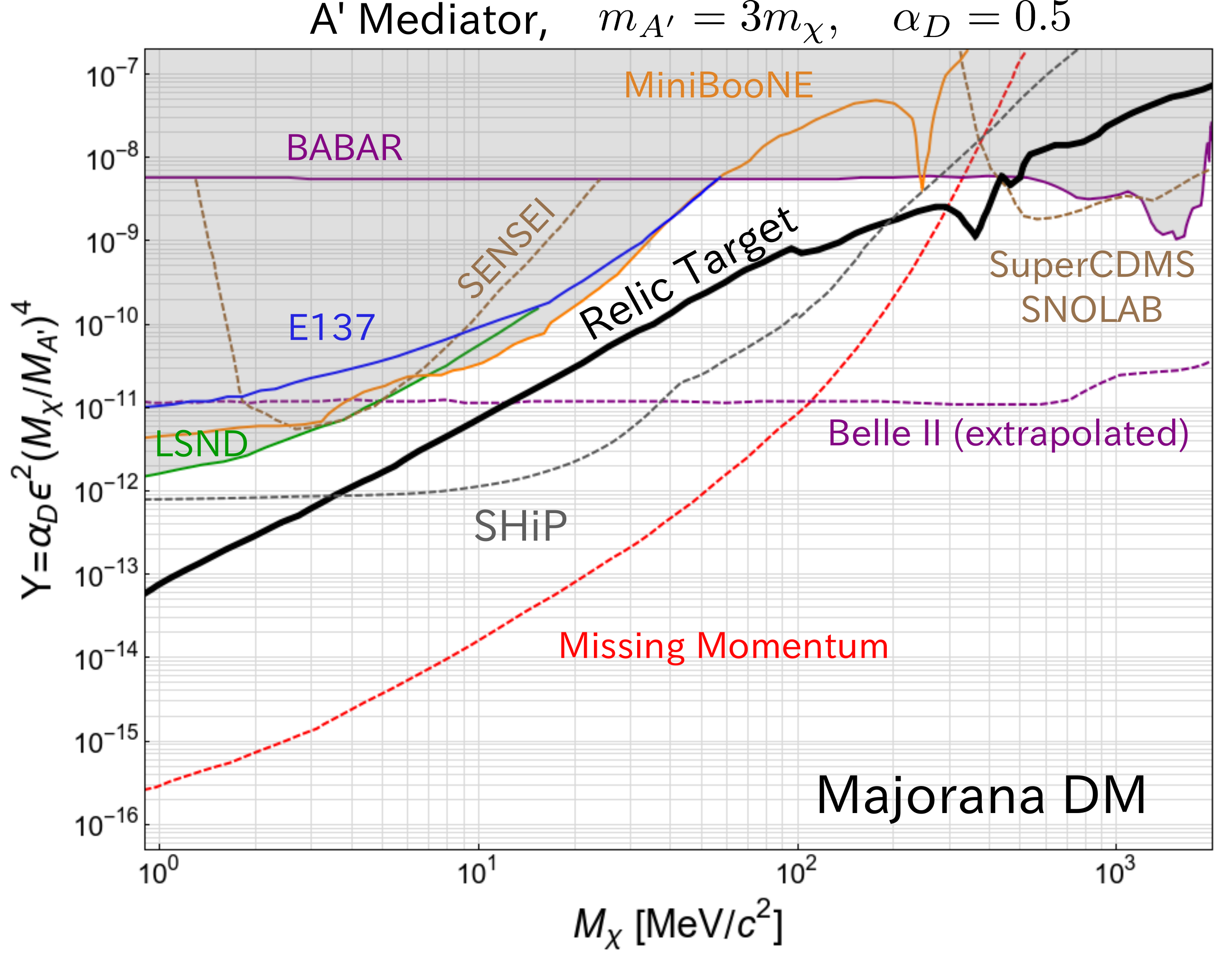}
\caption{\label{Fig:DM} \underline{Left}: Current (solid lines) and expected (dashed lines) constraints on the DM-nucleon 
scattering cross section from direct detection experiments~\cite{Bondarenko:2019vrb}.
\underline{Right}: Direct detection experiments results (solid lines) and projections (dashed lines) translated to 
the same parameter space as accelerator-based searches for Majorana DM particles~\cite{Akesson:2018vlm,Beacham:2019nyx}. Here the parameter $y$ characterizes both the DM abundance in the early Universe, and the DM-nucleon scattering cross-section. On the $x$-axis $m_\chi$ is the DM mass. The DM particles $\chi$ are assumed to interact with the SM via vector mediator $A'$.}
\end{figure}

A pioneer in direct detection light DM searches, the CRESST-III low-mass DM experiment~\cite{Abdelhameed:2019hmk} 
recently published its first result for DM masses below 1~GeV. It is about ten orders of magnitude weaker than the XENON1T results for DM masses of 10~GeV or higher (see Fig.~\ref{Fig:DM} left).
The SuperCDMS SNOLAB experiment has the potential to improve these constraints by about four orders of magnitude
in the future~\cite{Agnese:2016cpb}, but its potential impact on the DM parameter space strongly 
depends on the DM nature and other parameters, as demonstrated in Refs.~\cite{Akesson:2018vlm,Beacham:2019nyx} (see Fig.~\ref{Fig:DM} right). 

Despite these foreseen improvements, a large part of the  FIP parameter space will remain unexplored.
However, in the models with light mediators  the limited sensitivity of direct detection experiments can be mitigated by the searches of high-energy, beam-induced FIP particles.

LHC experiments, such as ATLAS, CMS or LHCb have limited sensitivities towards such searches.
Indeed, the visible decay modes of a light DM mediator are either outside the detector due to the long lifetime, or suffer from very large backgrounds in the low-mass mediator region that is relevant from the relic abundance point of view. 
Invisible decays of light DM typically produce quite low missing-momentum, as the dominating production mechanisms do not give a significant boost to the DM particles. Additional handles such as associated hard jets or vector boson production lead to increased model dependence and significantly suppress the cross section that can be probed. 

The most competitive results for light FIP searches therefore come from experiments based on electron beams that are able to control their backgrounds and precisely measure the energies of the final state particles in the events. 
These would include $b$-factories, such as BaBar and Belle in the past or Belle II~\cite{Kou:2018nap} in the future,
which employ the missing-mass technique; 
missing energy experiments at fixed target facilities, such as NA64~\cite{NA64:2019imj} operating at the CERN SPS; or the proposed LDMX experiment~\cite{Akesson:2018vlm};
and, finally, FIP scattering experiments to be operated at high-intensity proton beams, 
such as the future SHiP facility~\cite{Anelli:2015pba,Alekhin:2015byh}. 

The physics potential of a neutrino detector at LHC was discussed in a recent paper~\cite{Beni:2019gxv} where the TI18 tunnel was also identified as a suitable site to perform these measurements due to the very low  machine-induced background. A fully passive detector based on the emulsion technology was also recently proposed~\cite{Beni:2019pyp}. 

Here we propose to install a prototype of the Scattering and Neutrino Detector (SND) of the SHiP experiment at the LHC and collect data during Run 3 of LHC.  The detector will measure the neutrino energy and identify all three neutrino species in an unprecedented energy domain (between 350~GeV and few TeV). Given the pseudo-rapidity range accessible in this cavern, the detector will be off-axis and the corresponding neutrinos will mostly come from charm decays: the proposed experiment will thus make the first test of the heavy flavour production in a high pseudo-rapidity range that is not accessible by the current LHC detectors. 
The identification of the three species and the measurement of the energy spectrum will enable the muon and electron neutrino contributions to be disentangled, and a first ever estimate to be made for the heavy quark production in the forward region. This measurement will allow the study of quantum chromodynamics effects in an unexplored domain, which has significant implications for the simulation of heavy quarks produced in atmospheric swarms that are initiated by cosmic rays~\cite{Bhattacharya:2016jce}.
In addition, given the timing performance of the apparatus, we will be able to search for FIPs coming from the LHC. 
Finally, this will be an important test for the SND detector in view of its operation at the Beam Dump Facility.

\section{Overview of the experiment}

\begin{figure}[h!]
\centering
\includegraphics[width=0.45\linewidth]{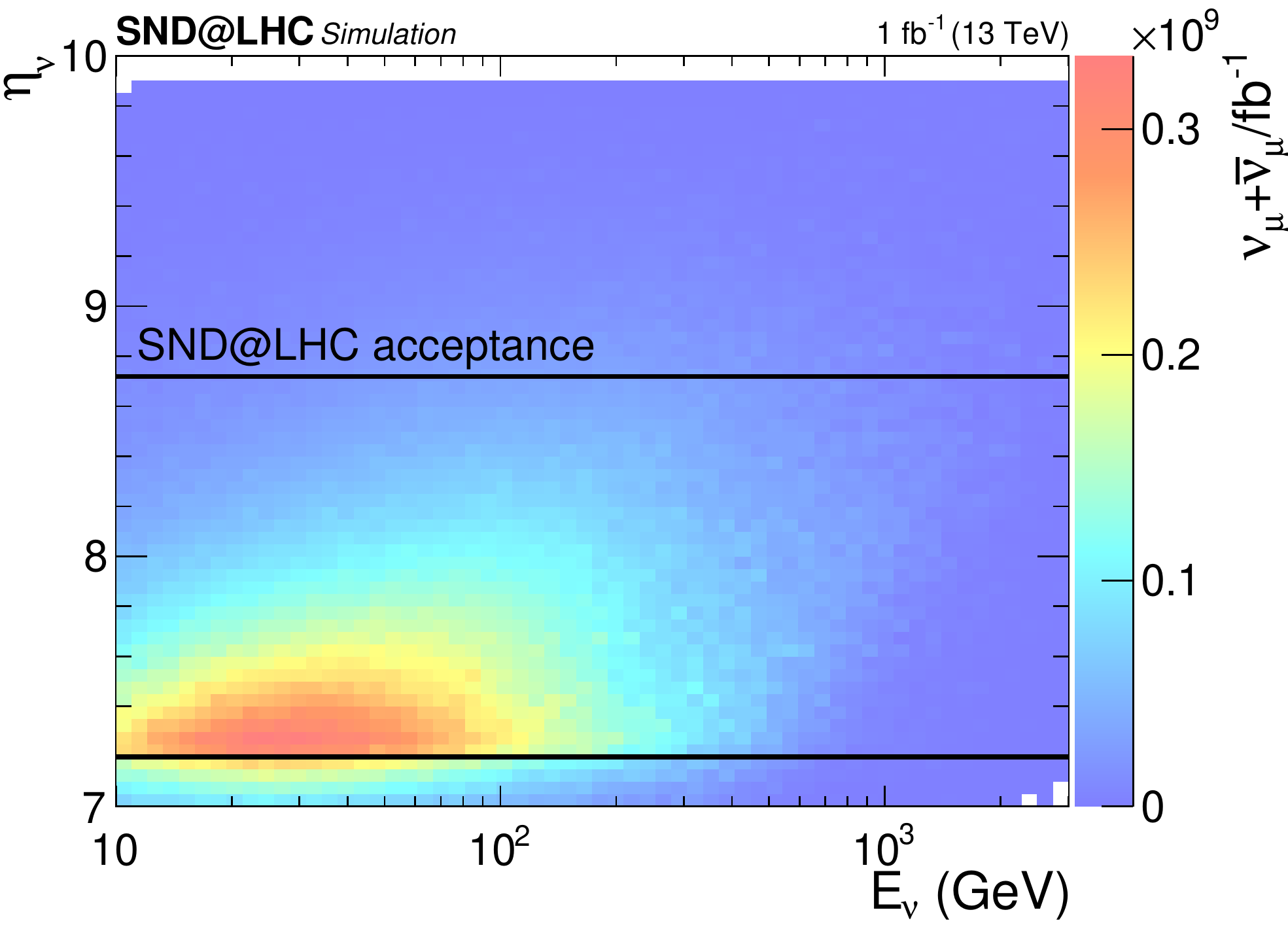}
\includegraphics[width=0.45\linewidth]{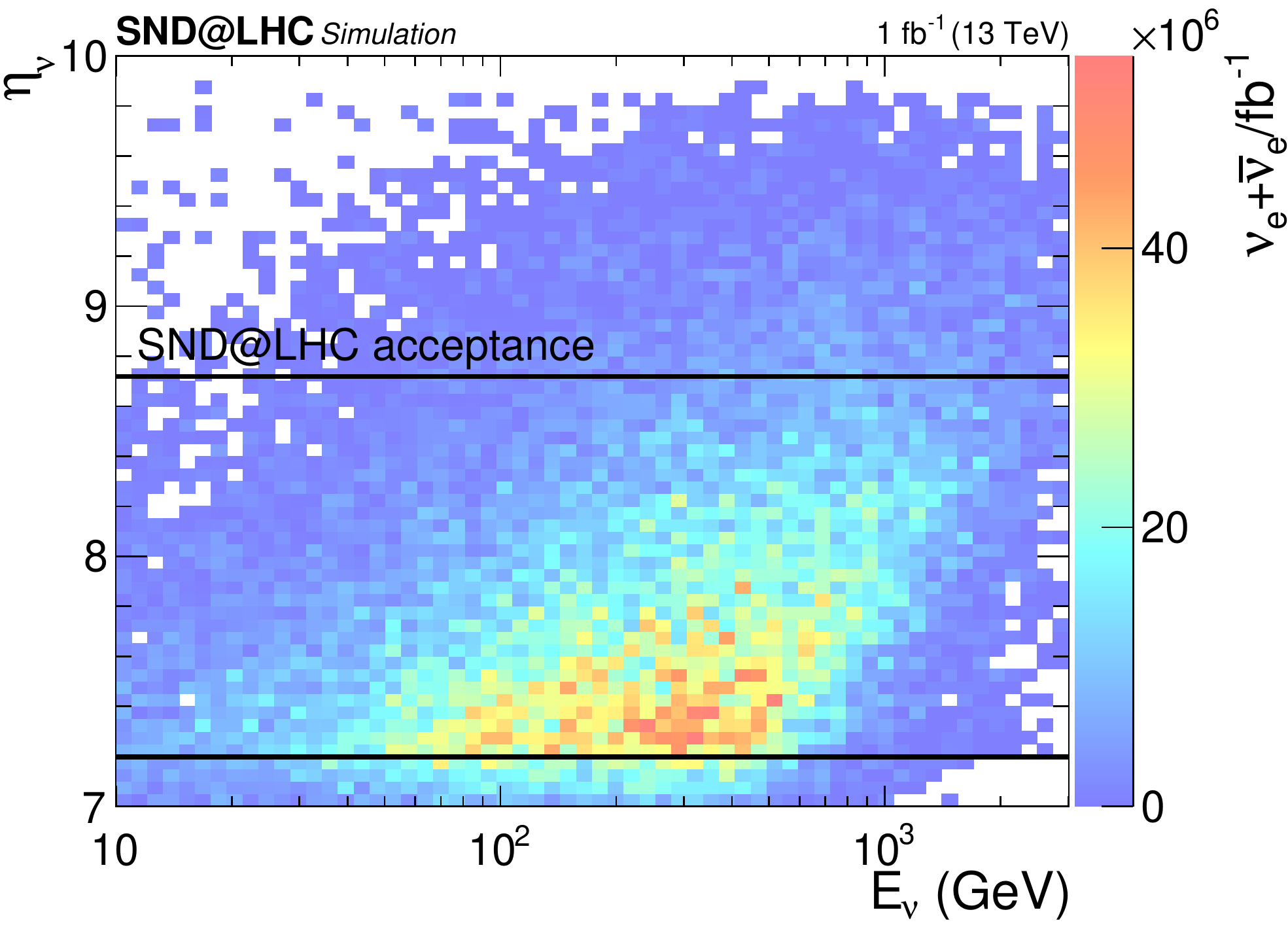}
\includegraphics[width=0.45\linewidth]{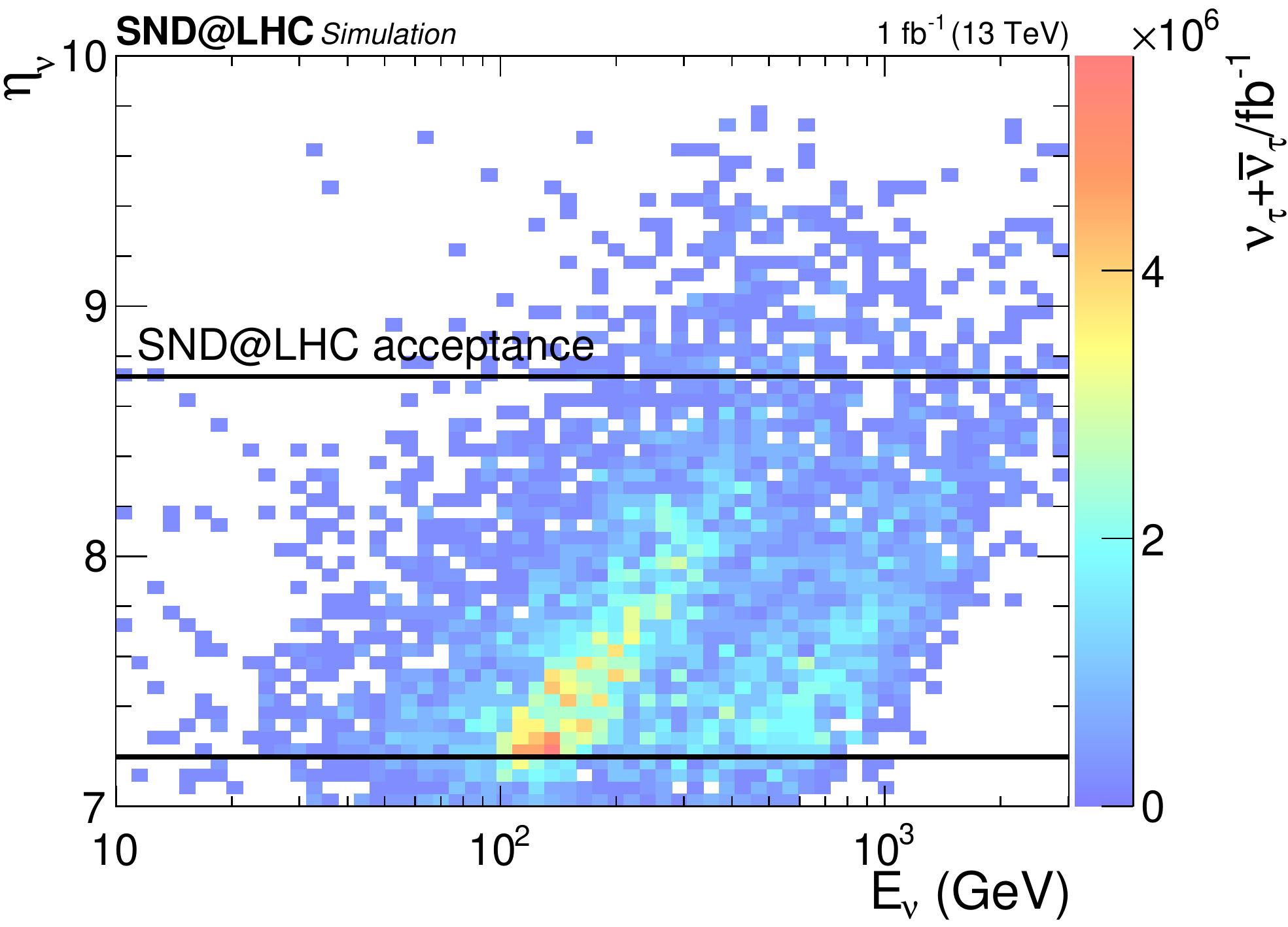}
\caption{\label{Fig:nuflux} 
Neutrino flux as a function of $\nu$ energy and pseudorapidity for muon (left), electron (right) and tau (bottom) neutrinos.}
\end{figure}

In the pseudo-rapidity range that will be covered, $7.2<\eta<8.7$,  electron and tau neutrinos originate practically only from heavy quarks, while the softer component of the muon neutrino flux is  also produced by pion and kaon decays. The flux of muon, electron, and tau neutrinos versus their pseudo-rapidity and energy is shown in Fig.~\ref{Fig:nuflux}. The heavy flavour production can be determined by measuring the yield of electron neutrinos. 
Assuming the SM cross-section for this species and the measured branching fractions for heavy  flavour to $\nu_{e}$, one can estimate the flux of the other neutrino species induced by heavy flavour production. 
This will allow the SM predictions to be tested and a search to be made for non-standard and flavour-specific neutrino interactions at high energies. 
In addition, the sensitivity to exotic scenarios, such as non-standard neutrino interactions and/or oscillations into light sterile states, will be studied.

The other  goal of the project is to search for light FIPs through their scattering off the atoms. Thanks to the ultra-fast timing, searches for FIPs producing nuclear recoil or higher multiplicity events will  be free of background in a large momentum and mass range. The sensitivity studies are yet to be completed, but significant improvements can be expected, especially for models with Dark Scalar mediators interacting predominantly with nuclei.

\begin{figure}[htbp]
\centering
\includegraphics[width=0.7\columnwidth]{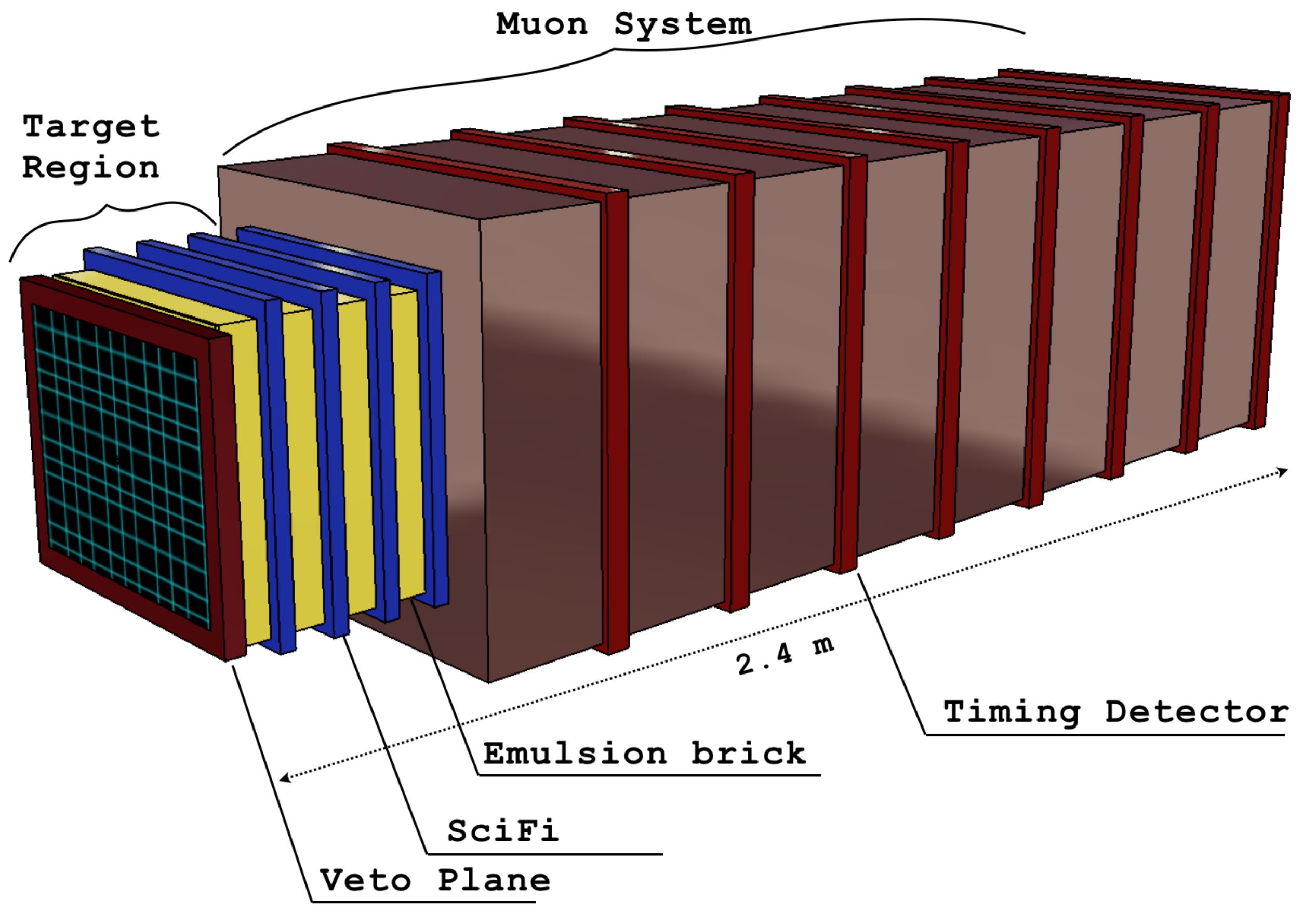}
\caption{Schematic view of the neutrino detector in its initial configuration.}
\label{layout}
\end{figure}

The challenging task of identifying the three neutrino flavours and FIP scattering  requires the construction of a hybrid detector, which combines nuclear emulsion technology and electronic detectors. A schematic view of the proposed detector is shown in  Fig.~\ref{layout}. 
The detector consists of a target region followed by a muon identification system. The target region is instrumented with nuclear emulsions and Scintillating Fibre (SciFi) planes. Upstream of the target region a plane of scintillator bars will act as a veto for charged particles. The muon identification system is located downstream of the target. It will consist of iron slabs interleaved with planes of scintillating bars. 

The emulsion detector, designed according to the Emulsion Cloud Chamber (ECC) technology, will act as a vertex detector with micrometric resolution, measuring all the outgoing charged tracks. The SciFi detector will predict the location of the neutrino interaction in the emulsion brick and will complement emulsion for the calorimetric measurement of electromagnetic showers. SciFi will also connect the emulsion track with the muon candidate track identified by the muon detector, thus allowing for the identification of muon neutrino charge-current interactions. Moreover, the combination of SciFi and the scintillating bars of the muon detector will also act as a non-homogenous hadronic calorimeter for the measurement of the energy of the hadronic jet produced in the neutrino interaction and hence the neutrino energy. 

The detector will be installed in the TI18 underground cavern, which has been identified as a suitable location in terms of machine-induced background~\cite{Beni:2019gxv}.

In charged-current interactions, the identification of the neutrino flavour will be made by identifying the charged lepton produced at the primary vertex.
Electrons will be clearly separated from $\pi^0$'s thanks to the micrometric accuracy, which will enable photon conversions downstream of the neutrino interaction-vertex to be identified.
Muons will be identified by the electronic detectors as the most penetrating particles. Tau leptons will be identified topologically in the emulsion, through the  observation of the tau decay vertex, together with the absence of any electron or muon at the primary vertex, according to the technology developed by OPERA~\cite{Agafonova:2010dc,Agafonova:2018auq}. 
FIPs will be identified through their scattering with the atoms of the emulsion target material. Both the scattering from  electrons and from nuclei will be exploited. 

The data taking will consist of a first run in 2021-2022 followed by a longer run, all along the LHC Run3. During the first run, in order to have more flexibility with the schedule, the detector installation will follow an incremental approach, reaching the detector configuration shown in Fig.~\ref{layout} by the end of 2021. A target mass of about 380 kg  will be installed, covering the pseudorapidity range $7.2<\eta<8.7$. The target will consist of four walls, each instrumented with 12 OPERA-like bricks (with 1 mm thick lead plates) and followed by a $40 \times 40$~cm$^2$ SciFi plane. The muon detector will consist of eight iron walls, for a total of 8 interaction lengths, interleaved by the same number of planes made of scintillator bars. The planes of scintillator bars will have a timing resolution better than 100~ps for the time-of-flight measurements of particles from the ATLAS interaction point. The resolution in the time-of-flight measurement will thus be determined by the 200~ps due to the spatial distribution of the particle bunches. The granularity of the bars in the three most downstream planes will be increased in such a way to improve the efficiency of the isolation criteria for the identification of penetrating muons. To this aim, the three most downstream planes will also be equipped with an additional plane with bars oriented in the orthogonal direction. 

During the technical stop at the end of 2022, the setup will be upgraded with an increase in the number of target walls, from 4 to 9 and the use of tungsten instead of lead, covering  the same pseudorapidity range. The target mass is increased to about 850 kg. Each wall will have the same thickness as for the lead target, i.e.~10~$X_0$, but would be thinner due to the shorter radiation length of the tungsten. The target region will amount to 90 $X_0$ rather than 40 $X_0$ as for the pilot run. Including the muon system, the total number of interaction lengths will exceed 11, thus improving the calorimetric performance. The three most downstream planes of scintillating bars of the muon system will be replaced by tiles to further increase the granularity, thus improving the muon identification efficiency.  During this operation, we foresee the replacement of the emulsion target every 25 fb$^{-1}$. This replacement can be performed within the technical stops, since it requires a short access.

 The muon system downstream will filter out hadrons;  muons will be left as the only penetrating particles. The hits of the muon system and those of the SciFi tracking station will have to be combined for the energy measurement of the hadronic shower induced by a neutrino interaction. 

Measurements already performed~\cite{Beni:2019gxv} foresee an expected charged particle rate in Run~3 lower than  $\rm 0.1\,Hz/cm^2$. On the other side, the detector will have to be protected from the thermal neutron flux with a dedicated shield. Given the expected neutron fluence of $\rm 4 \times 10^6/cm^2 /fb^{-1}$~\cite{Beni:2019gxv}, a few centimetres of boron carbide will be sufficient.  
\section{Experiment conceptual design}
\begin{figure}[htbp]
\centering
\includegraphics[width=0.8\columnwidth]{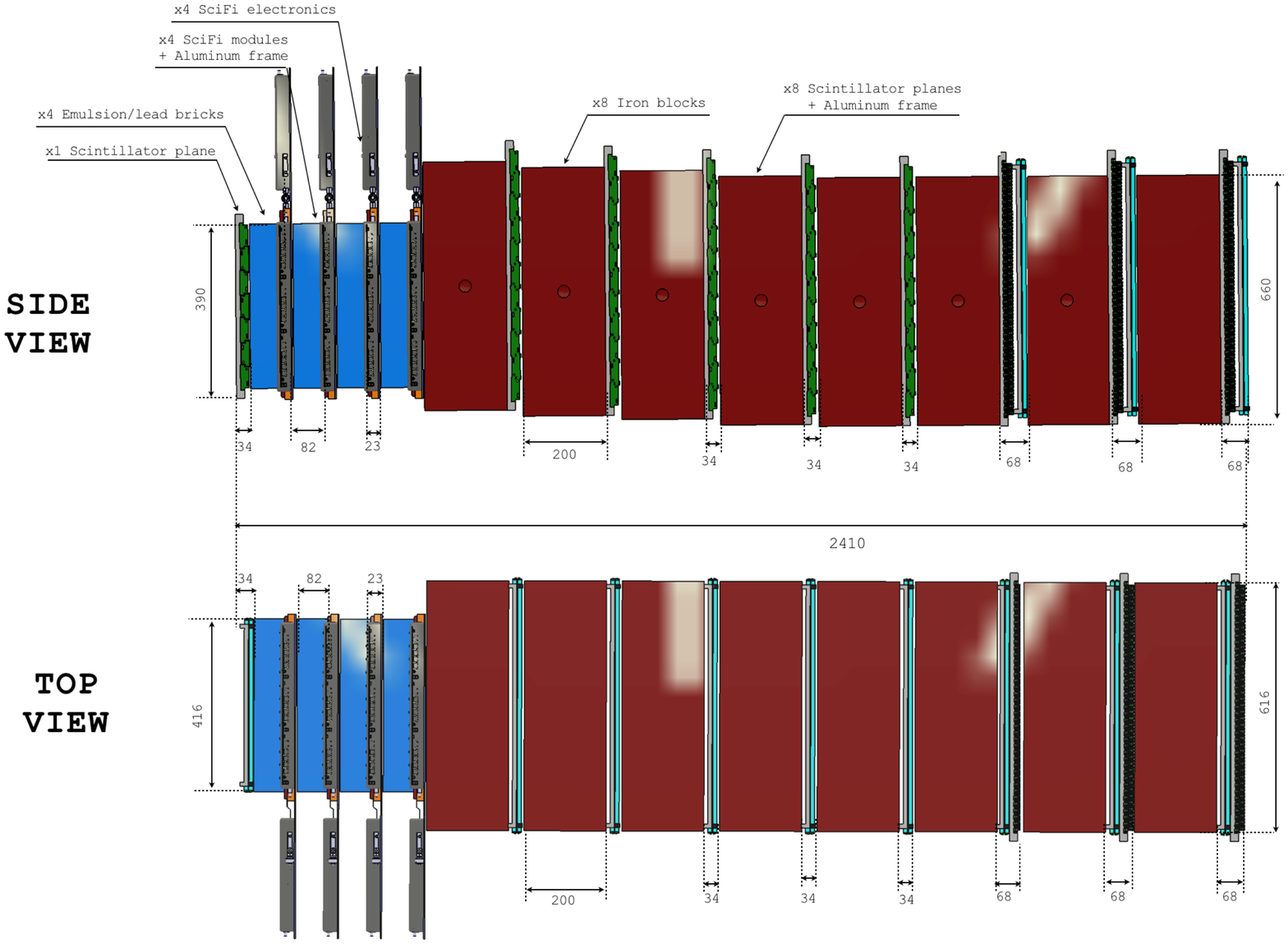}
\caption{Overall setup of the SND@LHC detector.}
\label{fig:overall_design}
\end{figure}

\begin{figure}[htbp]
\centering
\includegraphics[width=0.5\columnwidth]{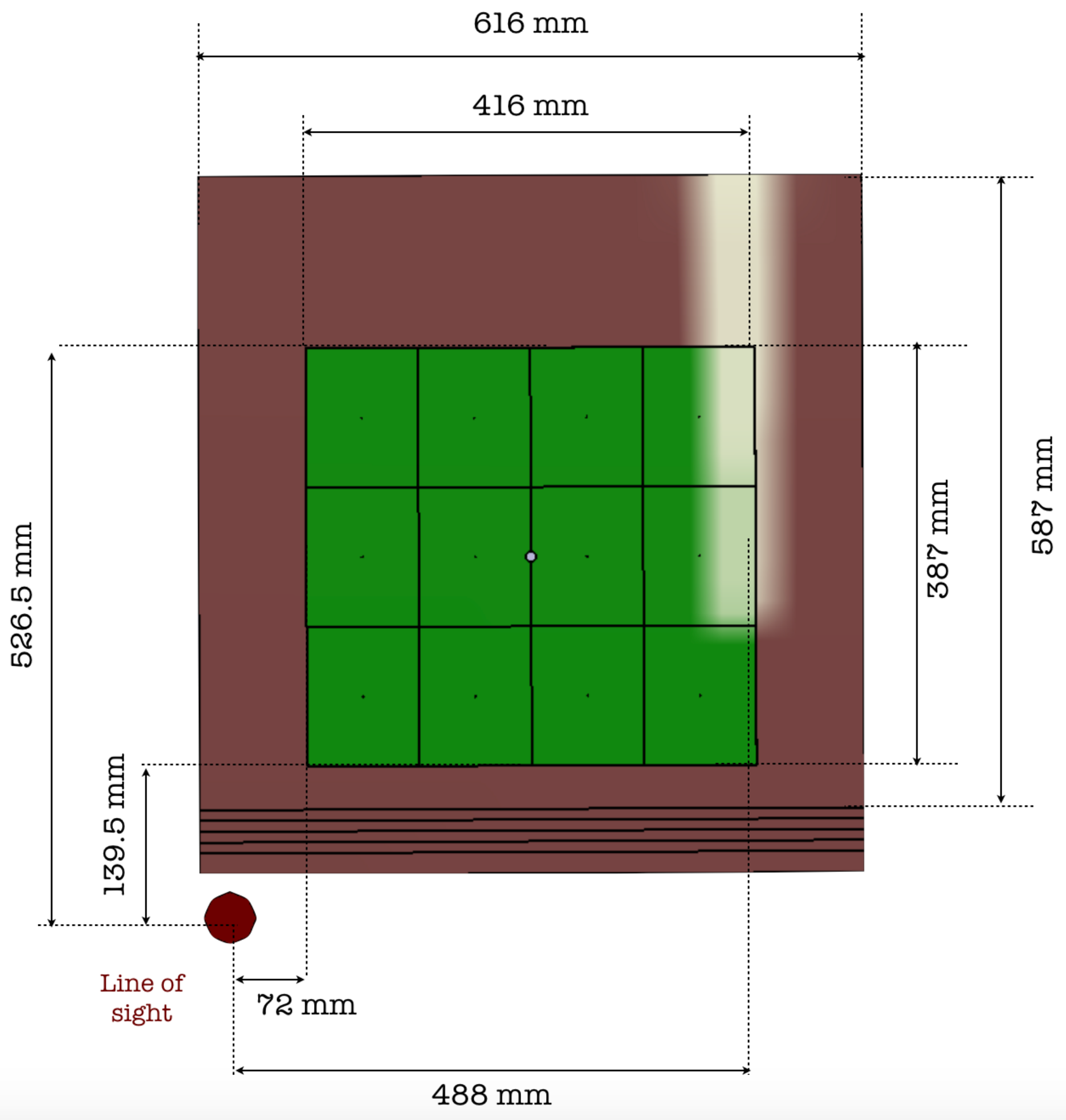}
\caption{Frontal view of the detector.}
\label{fig:layout_frontal}
\end{figure}

Figure~\ref{fig:overall_design} shows the design of the detector in its first run configuration. Both the side and top views are shown. 
Notice that the floor is 
inclined as it can be seen in the side view. 
For the target region, the electronics of the SciFi detector is also illustrated.  Figure~\ref{fig:layout_frontal} shows the detector in the transverse plane. 

\subsection{Upstream veto}
\begin{figure}[htbp]
\centering
\includegraphics[height=7cm]{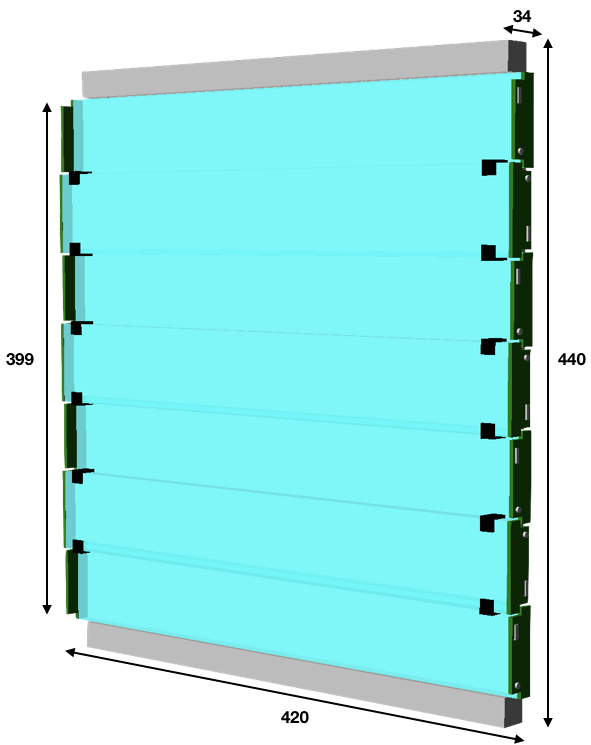}
\caption{Illustration of the Upstream Veto Detector. Units of the labels are in mm.}
\label{fig:veto_det}
\end{figure}

The Upstream Veto Detector, shown in Figure \ref{fig:veto_det}, will act as a veto for charged particles and will be located upstream of the Emuslion/SciFi detector. The baseline technology for the Upstream Veto Detector is scintillating bars read out by silicon photomultipliers (SiPM). The detector will cover an active area of 42 $\times$ 39.9 cm$^2$. It will comprise seven vertically staggered EJ200 plastic scintillating bars, each with dimensions of 42 $\times$ 6 $\times$ 1 cm$^3$ with  a 3 mm overlap between bars. Light generated in the bars by traversing particles will be collected and read out by an array of eight SiPMs on both ends. Each array of SiPMs will be read out by a custom made pre-amplifier PCB and the analogue signals subsequently digitised by a SAMPIC ASIC \cite{delagnes:in2p3-01082061}.

The material for the scintillating plastic was chosen by the timing resolution requirements. EJ200 is found to have the right combination of light output, attenuation length (3.8~m) and fast timing (rise time of 0.9~ns and decay time of 2.1~ns). The wavelength emission spectrum peaks at 425~nm, closely matching the SiPMs spectral response. The number of photons generated by a minimum-ionising particle crossing 1~cm of scintillating material is $\mathcal{O}(10^4)$. The bars will be wrapped in an aluminum foil and a black plastic stretch film on top to ensure opacity. 

\begin{figure}[htbp]
\centering
\includegraphics[height=7cm]{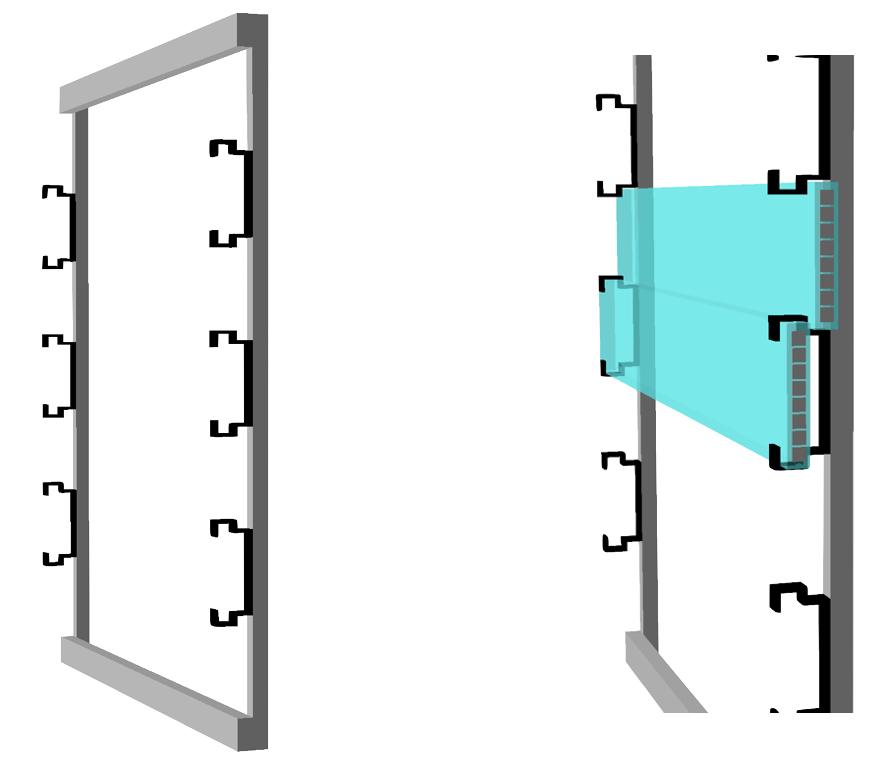}
\caption{Left:Aluminium frame of the veto detector with plastic supports for the scintillating bars. Right: Illustration of how the scintillating bars will be supported on the aluminium frame. The location of the eight SiPMs on the end of the bars is also shown.}
\label{fig:veto_support}
\end{figure}

An aluminium support frame will consist of two vertical columns supported on top and bottom by horizontal bars. Plastic supports for the scintillating bars will be attached to the vertical columns. The entire support frame can be seen on the left side of Figure \ref{fig:veto_support} and an example of how the scintillating bars will be fixed to the support is shown on the right side of Figure \ref{fig:veto_support}.

\begin{figure}[htbp]
\centering
\includegraphics[height=6cm]{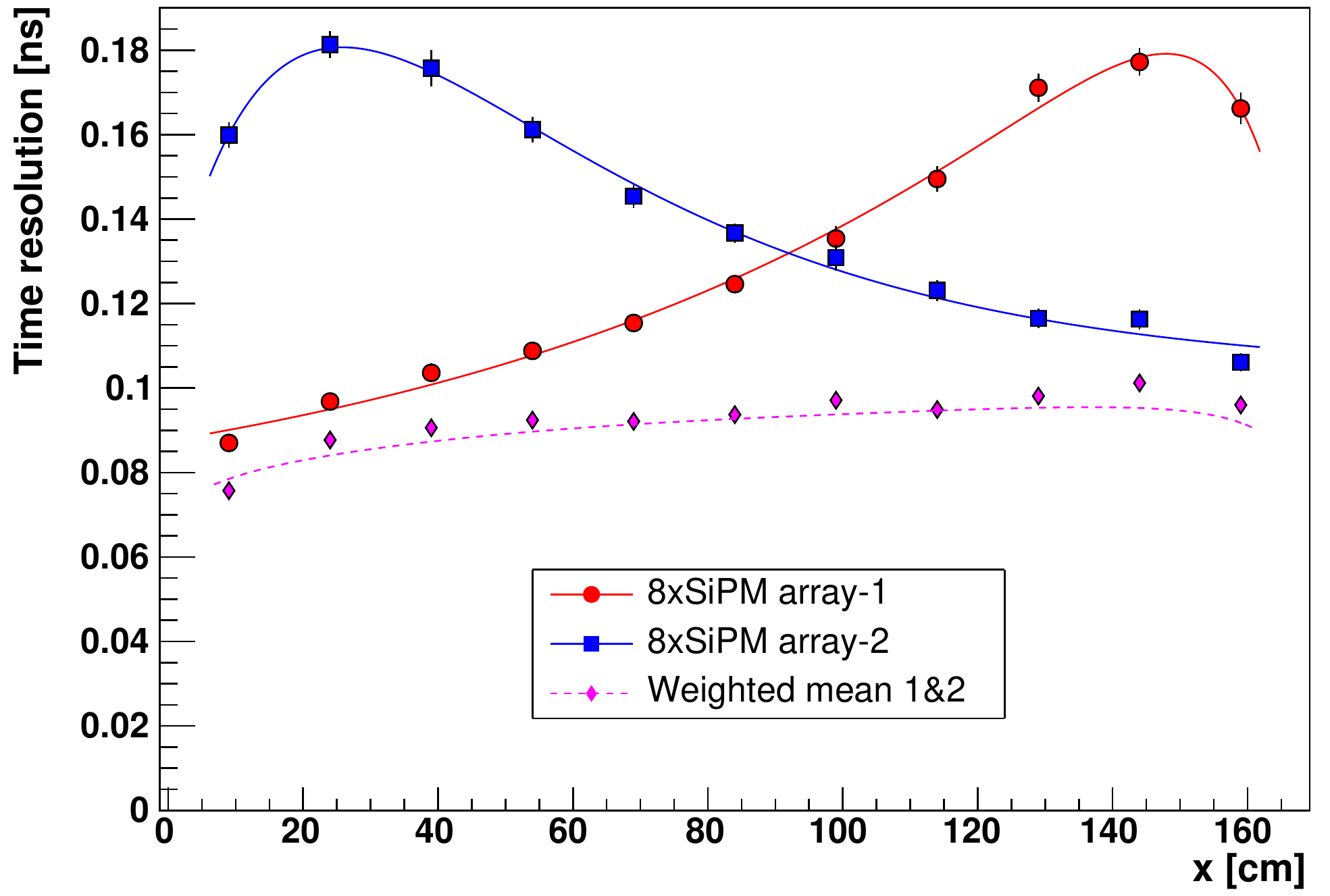}
\caption{Time resolution as measured by the SiPM arrays at both ends of a 1.68~m bar as a function of the beam impact position along the bar \cite{Korzenev:2019kud}.}
\label{fig:time_res}
\end{figure}

This technology has already been shown to act as a timing detector with resolution in the 10-100 ps range \cite{betancourt:2017,mu3e}. The design is based on the SHiP timing detector, where bars of 168 $\times$ 6 $\times$ 1 cm$^3$ are  read out on both ends by arrays of SiPMs. A 22 bar prototype of this detector was built and tested in summer 2018 and a timing resolution of 85 ps was achieved using the weighted mean from both ends of a single bar, as seen in Figure \ref{fig:time_res}. Extrapolating these results to the length of the bars for the upstream veto detector gives an expected timing resolution of $\sim$ 50 ps. Our requirement of 100 ps is therefore conservative. 

\subsection{Emulsion target}
\label{subsec:ECC}
\begin{figure}[htbp]
\centering
\includegraphics[width=0.8\columnwidth]{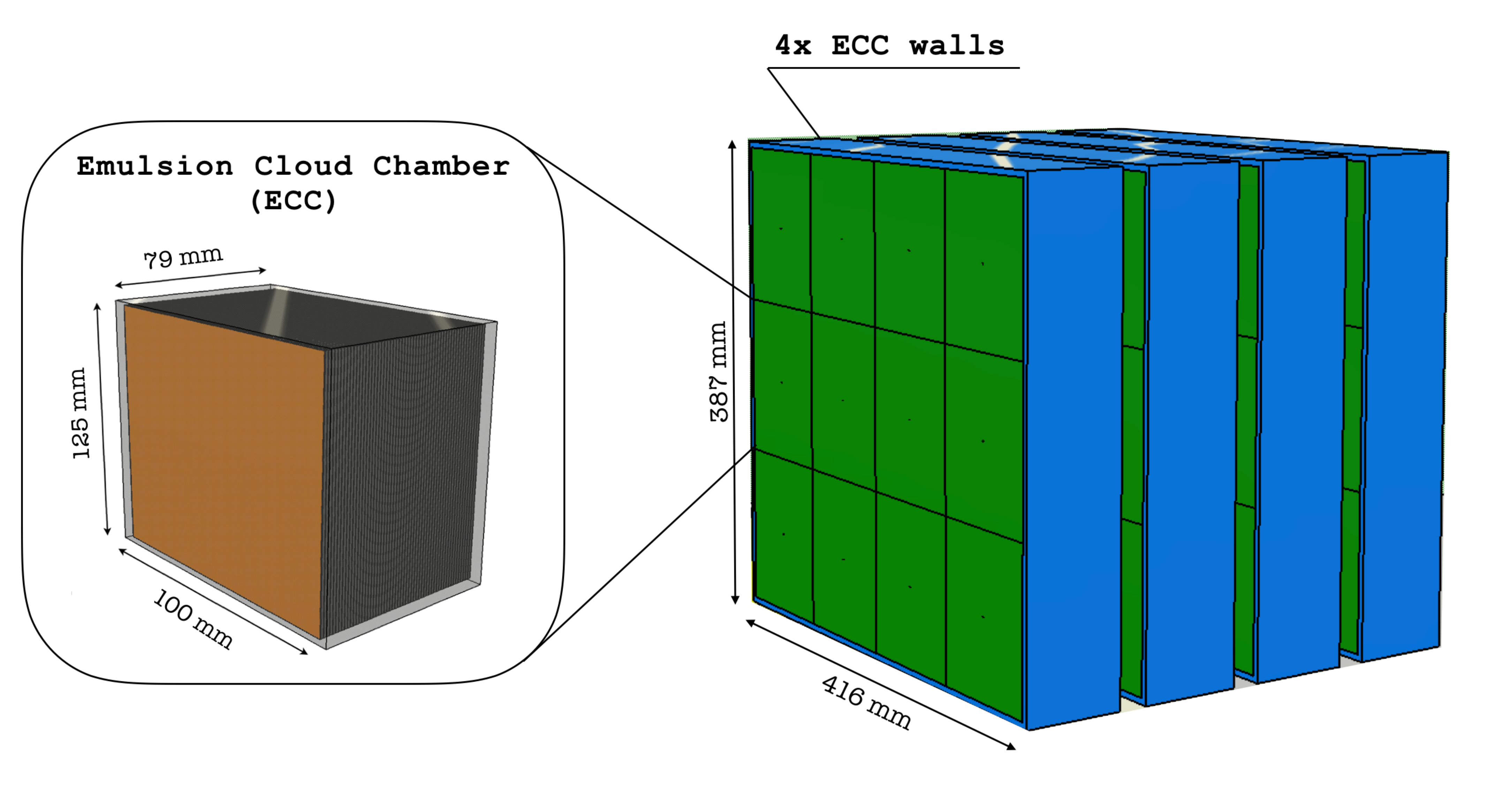}
\caption{Layout of the emulsion target, consisting of four walls, each made of 12 ECC units.}
\label{fig:emulsion_target}
\end{figure}

The Emulsion target is made of four emulsion brick walls and four Target Tracker planes. The walls are divided in 3$\times$4 cells, each with a transverse size of 100$\times$125 mm$^2$, consisting of Emulsion Cloud Chambers (ECCs) as illustrated in Figure~\ref{fig:emulsion_target}.

The ECC technology makes use of nuclear emulsion films interleaved with passive absorber layers to build up a tracking device 
with sub-micrometric position and milliradian angular resolution, as demonstrated by the OPERA experiment~\cite{Acquafredda:2009zz}. 
It is capable of detecting $\tau$ leptons~\cite{Agafonova:2018auq} and charmed hadrons~\cite{Agafonova:2014khd} by disentangling their production and decay vertexes. It is also suited for FIP detection through the direct observation of the scattering off electrons in the absorber planes. The high spatial resolution of  nuclear emulsion films allows  identifying electrons by observing electromagnetic showers in the brick~\cite{Agafonova:2018dkb}.
Nuclear emulsion films are produced by Nagoya University  and by the Slavich Company in Russia. For the long-term stability of the emulsion films, the temperature of the target will be  kept at 15$^\circ$C. 

A unit cell is made of 57 emulsion films with a transverse size of  125$\times$100 mm$^2$, interleaved with 1\,mm thick lead layers. The resulting brick has a total thickness of $\sim$8 cm, corresponding to $\sim$10$X_0$, and a total weight of $\sim$8\,kg. The overall target weight with four walls of 4$\times$3 bricks is about 380\,kg. The films are analysed by fully automated optical microscopes~\cite{Arrabito:2006rv,Armenise:2005yh}. The scanning speed, measured in terms of film surface per unit time, was significantly increased in recent years~\cite{Alexandrov:2015kzs,Alexandrov:2016tyi,Alexandrov:2017qpw}. R\&D is still ongoing~\cite{Alexandrov:2019dvd} to further increase the scanning speed.  

\begin{figure}[h!]
\centering
\includegraphics[width=1.0\linewidth]{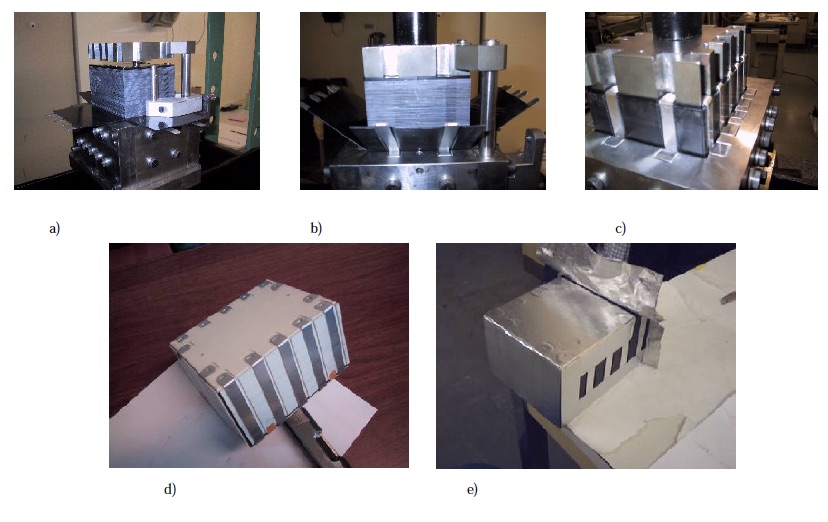}
\caption{Sequence of the spider packaging procedure.}\label{fig:spider_procedure}
\end{figure}

\begin{figure}[h!]
\includegraphics[height=0.3\linewidth]{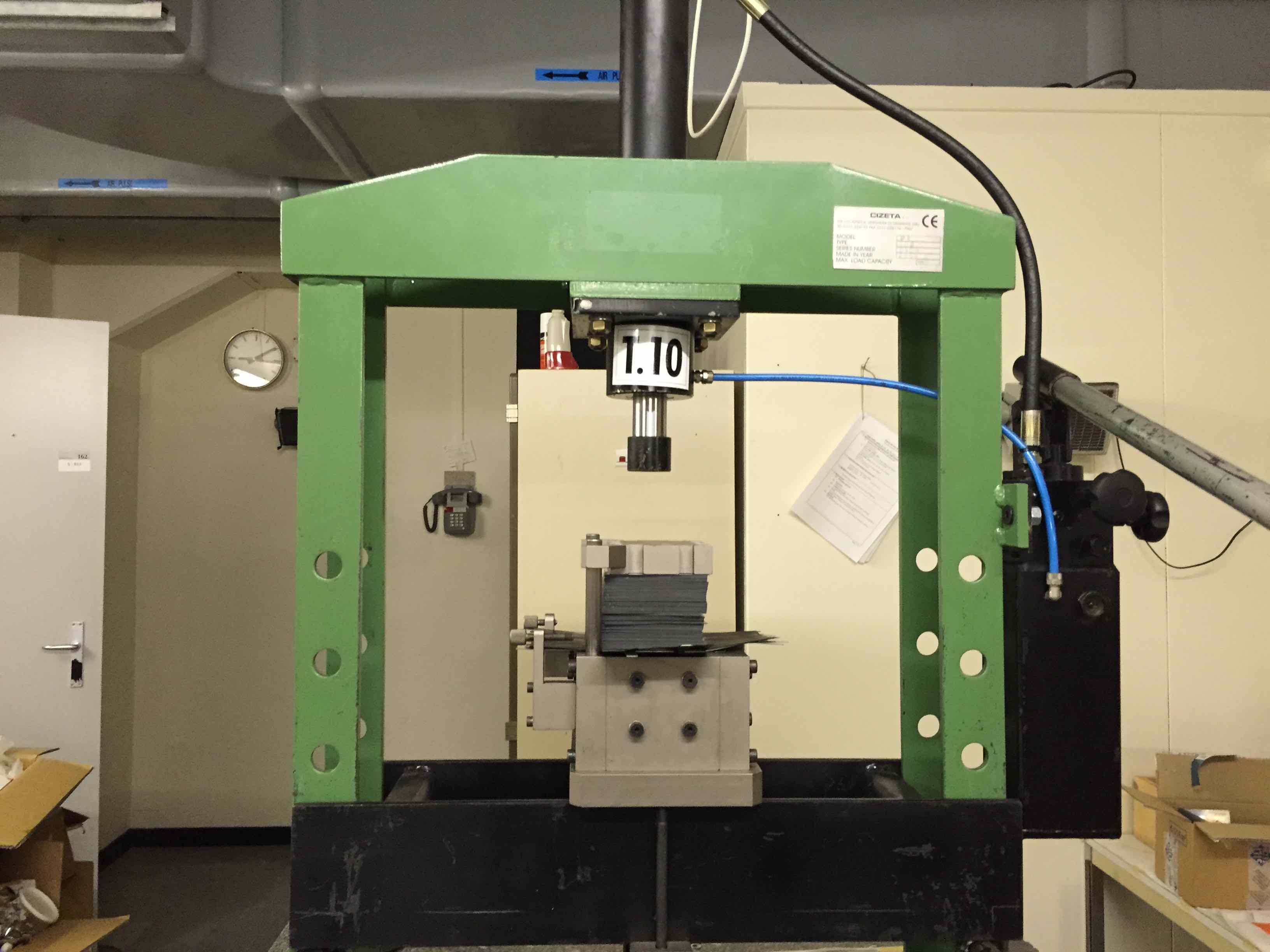}
\includegraphics[height=0.3\linewidth]{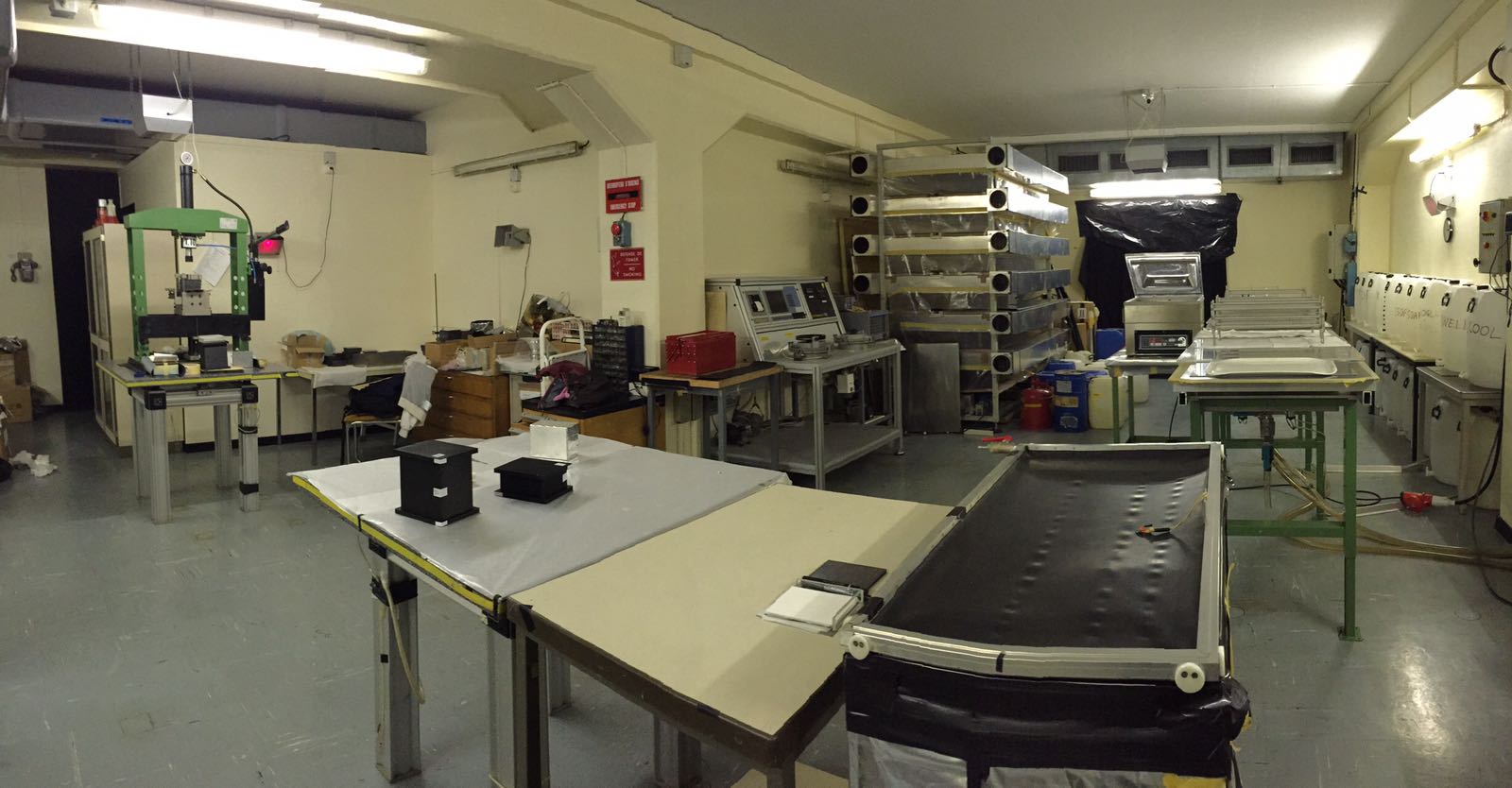}
\caption{Left: press used in the emulsion laboratory at CERN to prepare the ECC target. Right: the SHiP emulsion laboratory at CERN.}\label{fig:emu_lab}
\end{figure}

\subsubsection{Assembly procedure}
We propose to assemble  emulsion films and passive layers using the packaging procedure adopted in the OPERA experiment and commonly referred to as `spider packaging procedure'. It is based on a 800~$\mu$m thin aluminum foil, called `spider', that provides mechanical stability to emulsion films and passive layers, which are stacked together to form a pile. The spider is firstly placed under the pile (Figure \ref{fig:spider_procedure}a), then it is folded on the sides by mechanical pressure (Figure \ref{fig:spider_procedure}b) and closed on the upper emulsion film (Figure \ref{fig:spider_procedure}c). Plastic side protection and cover keep the rigidity and avoid the direct contact between emulsions and aluminum (Figure \ref{fig:spider_procedure}d), the light shielding is provided by wrapping an adhesive aluminum tape around the pile (Figure \ref{fig:spider_procedure}e). 

We plan to perform the preparation of the ECC target and subsequent development of emulsion films in the SHiP emulsion laboratory at CERN (see Fig.~\ref{fig:emu_lab} left). The required spider and the press are already available (Fig.~\ref{fig:emu_lab} right).

\subsection{Target Tracker}
\label{subsec:tt}
The scintillating fibre (SciFi) tracker technology is particularly well suited for large surface 
tracking in a low radiation environment and it will play a key role to deliver the physics goals of the experiment. 
It allows to disentangle piled up events in the emulsion and it provides both time and energy information 
to the tracks and showers and associate these to the recorded events in the emulsion films. 
The SciFi detector technology for the target tracker, fibre mats and photo detectors, were developed by the EPFL group for 
the LHCb tracker upgrade~\cite{LHCbCollaboration:2014tuj}.
While the SciFi provides tracking measurements for charged particles in the LHCb experiment, it serves 
several more purposes in the SND@LHC. 
Firstly, thanks to its spatial resolution of the order of 50 $\mu$m, it will connect  tracks reconstructed by 
the electronic detectors with the corresponding tracks seen in the emulsion, thus providing the time stamp 
to the particle interaction detected in the emulsion films. It will also act as an active part of a sampling 
calorimeter of 40 radiation lengths, $X_0$, depth made together with the ECC bricks. 
Moreover, the target region and the downstream muon system form together a hadronic calorimeter, 
with about 9.3 interaction lengths, $\lambda$, in the pilot run and more than 11~$\lambda$ in the extended run. 
Therefore the SciFi plays an important role in the energy measurement of the hadronic 
and electromagnetic jets induced by the incident particle interaction. 

\begin{figure}[h!]
\centering
\includegraphics[height=0.33\linewidth]{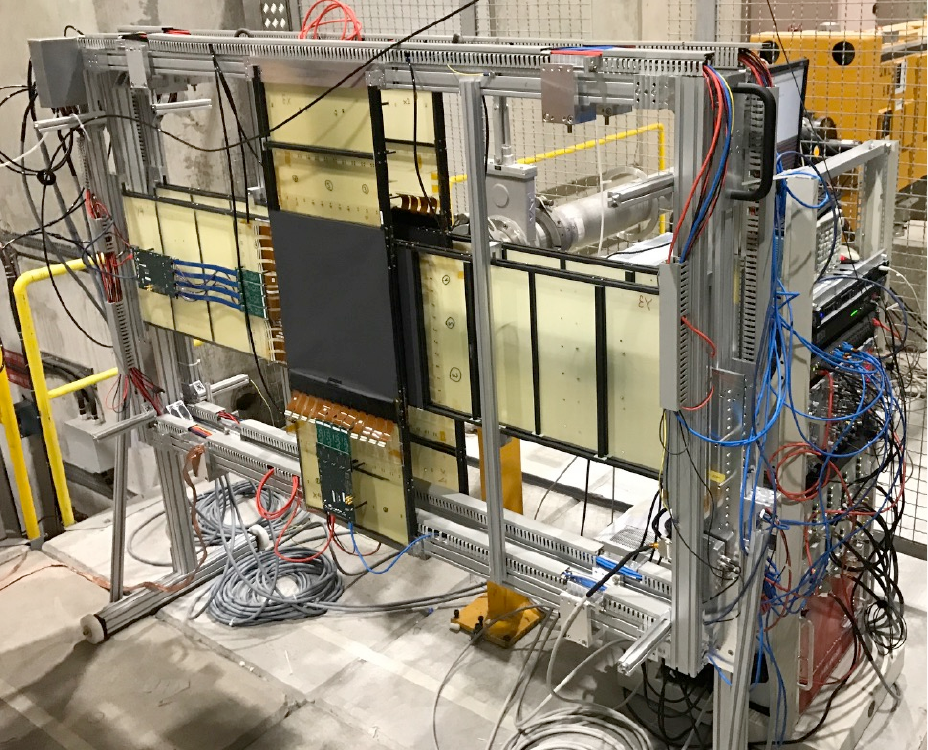}
\includegraphics[height=0.33\linewidth]{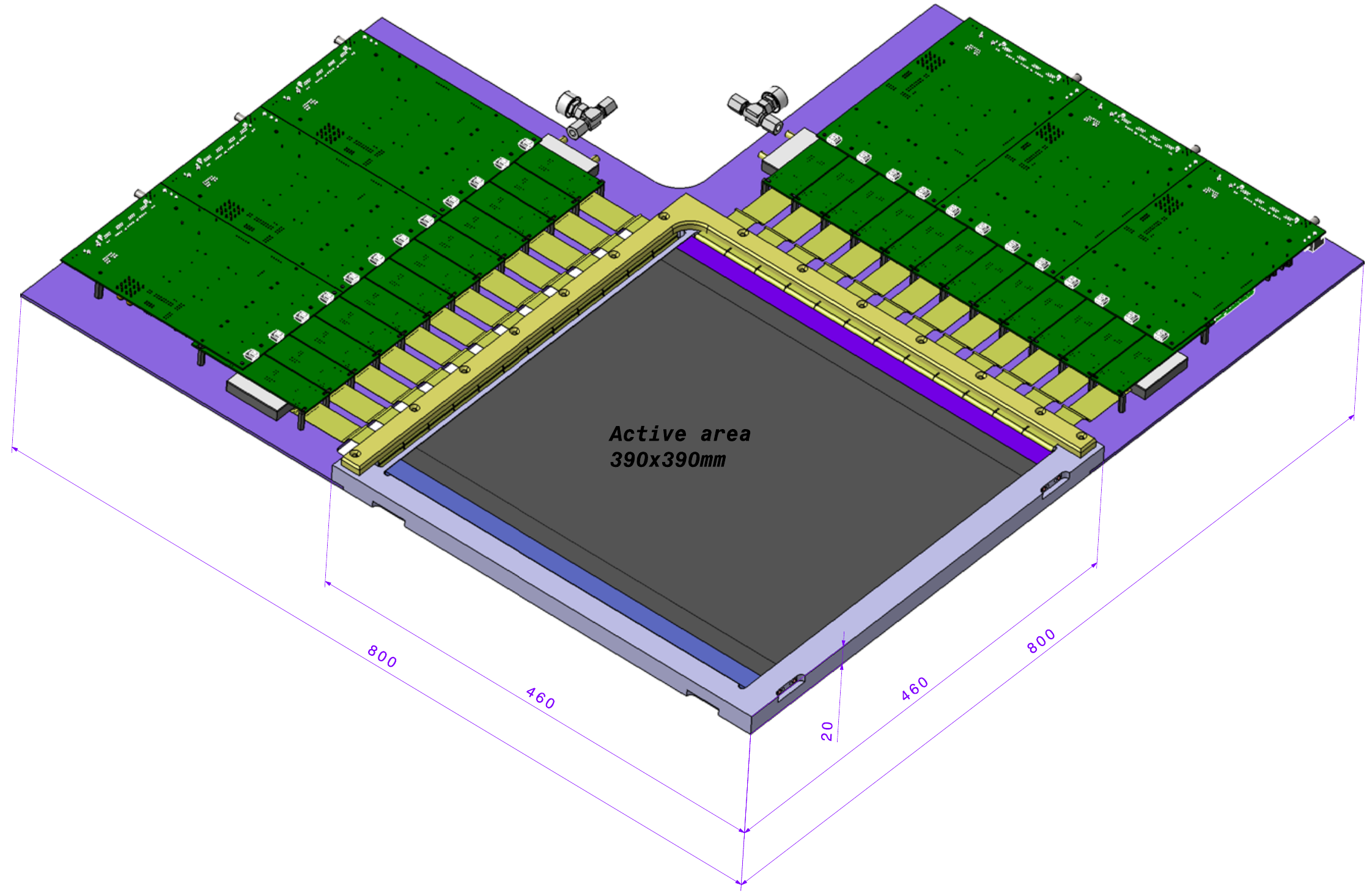}
\caption{\label{Fig:scifi_drawing} (left) SHiP-Charm SciFi modules equipped with readout. (right) A SciFi X-Y detection module with its corresponding readout electronics and the water cooling system for the front-end ASICs. }
\end{figure}
  
Four $x$-$y$ SciFi planes equipped with the readout electronics, have been produced for the SHIP-Charm experiment~\cite{Akmete:2286844},
and may be used in the SND@LHC pilot run in 2021 (Fig.~\ref{Fig:scifi_drawing} left).

The detection layers are based on the blue light emitting Kuraray SCSF-78MJ scintillating fibres of 250\,$\mu$m diameter with a decay time of 2.8\,ns.
The layers are made of six densely packed staggered fibre layers glued together forming fibre mats of 133\,mm width and 390\,mm length; three fibre mats are integrated to a fibre plane with less than 500\,$\mu$m dead zones between mats (Fig.~\ref{Fig:scifi_drawing} right). 
The readout channel segmentation with 250\,$\mu$m wide channels is provided by customised 128-channel SiPM array developed by the EPFL group with Hamamatsu for the LHCb experiment. A total of about 3000 channels is required for an X-Y module in the target tracker. 
The readout is situated outside the acceptance and consists of the photo-detector located at the fibre module's edge, 
a short Kapton flex PCB and the front-end electronics board. Light shielding of the assembly is ensured by 
a seal implemented at the flat Kapton flex section enclosing the photo-detectors and the entire fibre region. 
This part doesn't require any cooling as the heat dissipation of the SiPMs is sufficiently low. For time calibration and threshold adjustment, a light injection system is placed on the opposite side of the photo-detector fibre end.  
For the time measurement with the SciFi, a readout system based on the STiC ASIC~\cite{Harion_2014} 
has been developed and tested in several testbeam campaigns during 2018-2019. 
The coincidence time resolution (CTR) of 350~ps between two planes of the size 133\,mm$\times$133\,mm 
 has been demonstrated with  minimum ionizing particles. This corresponds
to a single plane time resolution of about 250\,ps. In case of the calorimetric operation of the SciFi,
which leads to an order of magnitude more photons produced in fibres, further improvement of the time 
resolution is expected.
An FPGA on the readout board is collecting the hit information of the STiC chips and it is assembling them into larger frames. The FPGA handles the interface between the front-end and the DAQ CPU over a Gigabit Ethernet for data acquisition and slow control.
The hit detection efficiency of the SciFi is expected to be larger than 99\% for large planes, since no radiation 
damage will occur during the operation of the detector. 
\subsection{Muon and timing detector}
\label{subsec:timdet}
 The muon and timing detecgor is located downstream of the ECC/SciFi detector and it will identify muons, crucial to identifying muon neutrino charged-current interactions. In combination with the SciFi, it will serve as a non-homogeneous hadronic calorimeter, enabling measurement of the energy of hadronic jets. The timing resolution of about 100 ps per hit will allow to timestamp tracks in the SciFi and emulsion by linking them to tracks in the muon system. 

\begin{figure}[htbp]
\centering
\includegraphics[height=7cm]{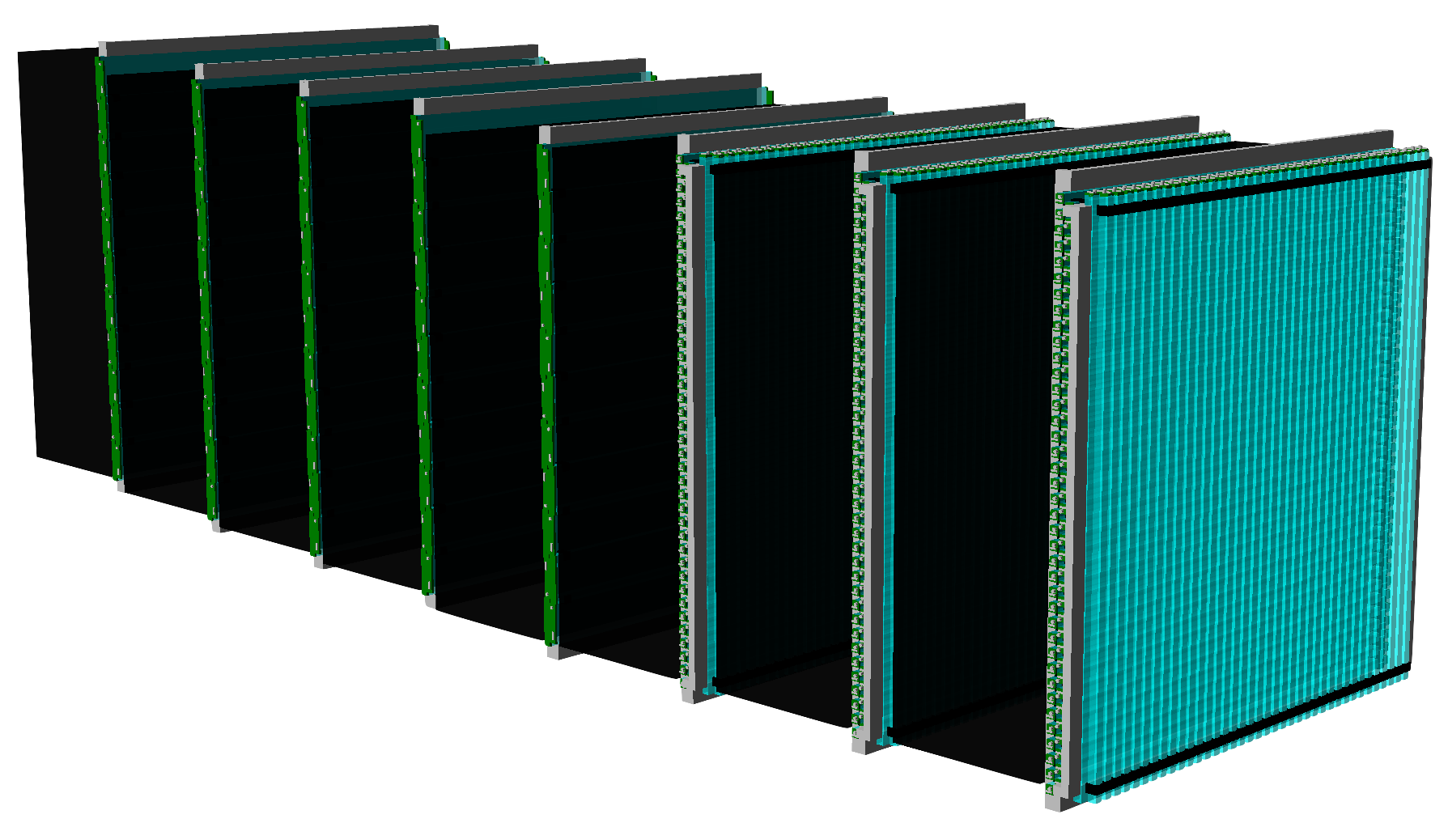}
\caption{A schematic representation of the muon and timing detector.}
\label{fig:muon_id}
\end{figure}

Eight scintillating planes will be interleaved between layers of iron slabs 20 cm thick, which will act as passive material. The first five planes are similar to the upstream veto detector, albeit with different dimensions, and will be used as a timing detector for traversing particles. The last three layers consist of two layers of thin bars, one arranged horizontally and one arranged vertically, allowing for a spatial resolution less than cm. Muons will be identified as being the most penetrating particles as described in section \ref{subsec:neutrino}. A schematic representation of the muon identification system in the pilot run is shown in Figure \ref{fig:muon_id}.

\begin{figure}[htbp]
\centering
\includegraphics[width=14cm]{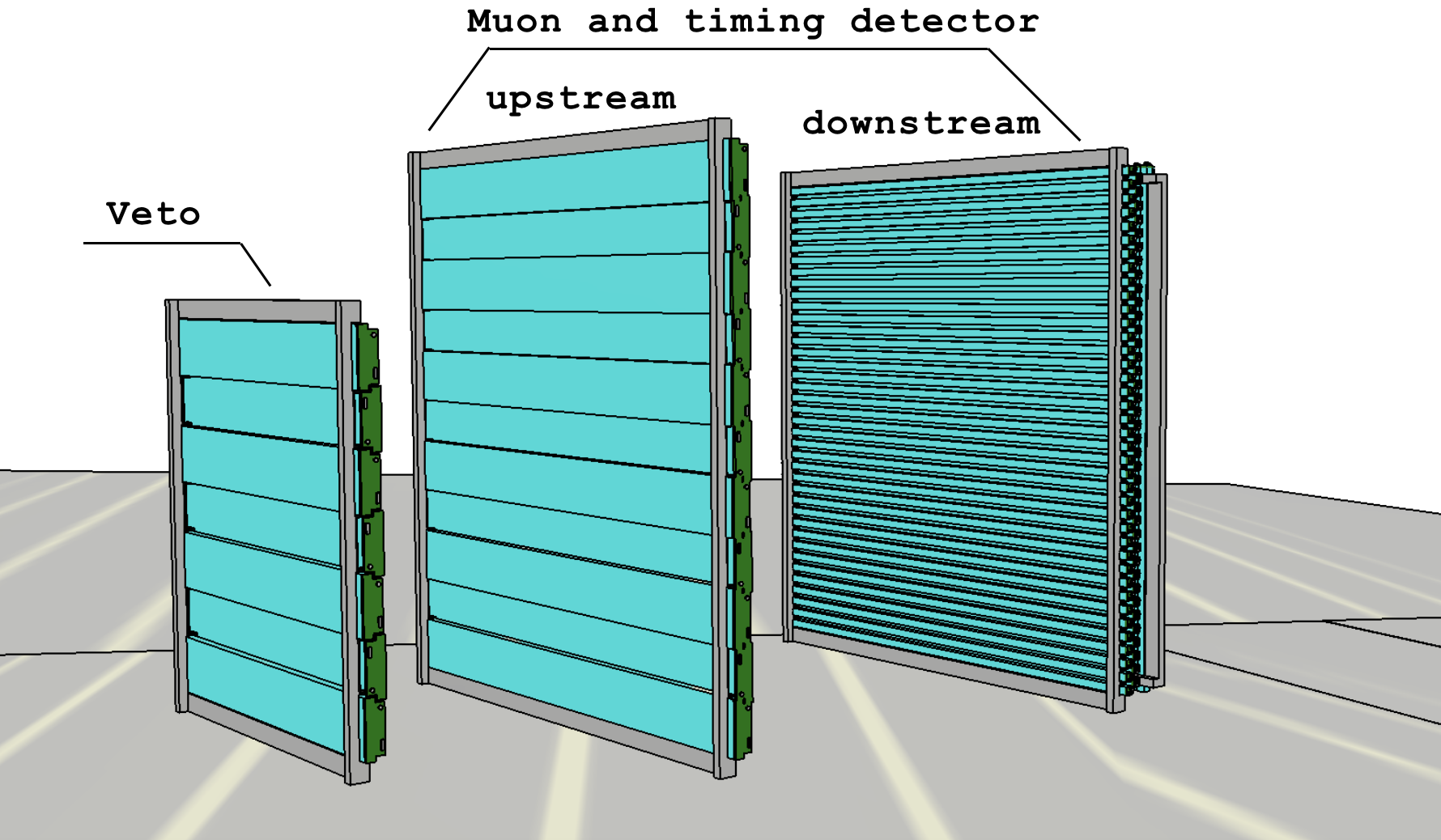}
\caption{Layout of the upstream and downstream configuration of the muon and timing detector planes. The veto plane is also shown on the left side for comparison. }
\label{fig:muon_timing}
\end{figure}

The five timing layers comprise 11 bars, each with dimensions 62 $\times$ 6 $\times$ 1 cm$^3$ with a 5 mm overlap between bars. Each layer covers an active area of 62 $\times$ 60.5 cm$^2$. Every bar is wrapped in an aluminum foil and a black plastic stretch film and read out by a custom pre-amp PCB housing eight SiPMs on both ends, similar in fashion to the upstream veto detector. The timing resolution of each timing layer is expected to only be slightly worse than the upstream veto detector. 

The last three layers will each consist of two scintillating planes; one with bars arranged horizontally and the other with vertically arranged bars. Bars in both directions measure 62 $\times$ 1 $\times$ 1 cm$^3$ with a 2 mm overlap between bars. The horizontal bars are read out on each end by a single SiPM, while the vertical bars are read out only from the top due to constraints of the cavern floor. 

An intensive R\&D period during Phase 1 will focus on replacing the bars in the last three layers with scintillating tiles. This will increase both the spatial and timing performance while also reducing the occupancy per channel, helping with the muon isolation and identification. The tile planes will then be installed at the end of 2022.

\subsection{Data acquisition}
The DAQ system close to the detector is composed of three parts, the fast control and timing signal distribution, two separate Gigabit Ethernet networks for the Timing detectors and the SciFi tracker readout and the slow control network for the infrastructure such as the cooling unit, temperature and radiation monitoring. The remote and strongly access restricted location as well as the power limitation at the TI18 cavern, suggests to locate the online data processing equipment on the surface. A possible location for a surface rack location where the online data processing and storage can be performed has been identified. 
\subsubsection{Acquisition mode}
The signal from neutrino or dark matter particles is present only in a fraction of the target region and therefore only in one or several SciFi planes, making a hardware trigger implementation impossible. A data driven DAQ without trigger is imposed. All detector data, noise and physics event data, is transferred to the DAQ processor followed by a online data analysis for noise rejection and event building.       
\subsubsection{Timing and synchronisation}
The synchronisation with the LHC machine will be implemented with the dedicated  timing, trigger and control (TTC) systems. The TTC signal distributed on a single mode optical fibre has to be routed to the experimental site. The synchronisation signals Clock, Trigger and Reset will be made available for the synchronisation between the LHC, the timing detector and the SciFi tracker. The time of flight of the detected particles can be extracted from the difference of the LHC clock and the particle hit time. The periodic injection of time calibration and synchronisation triggers will allow to verify the synchronisation of all electronics component over the long time run periods. For event building, all hit data is time stamped with a 32-bit long (LHC clock cycle unit,25\,ns) time counter representing a time interval of 107\,s. The same time counter is sent to the DAQ up on a calibration trigger and allows for periodic verification of synchronisation. 
\subsubsection{Data rates and event processing}
The data rates in the detectors are dominated by the noise data rate. For the timing detector the estimated data rate per 64-channel SAMPIC module is 150\,Mbit/s and for the system of the pilot run (13 modules) 2\,Gbit/s. For the SciFi tracker the expected hit noise rate per readout channel is 100\,Hz (64-bit per hit). The number of channels for one SciFi X-Y plane is 3\,K, with 4 planes the expected data rate is 80\,Mbit/s. The data streams from the different modules are aggregated on a Gigabit Ethernet switch located in the detector rack and transmitted over 10 Gigabit Ethernet optical links to the DAQ server rack located on the surface. For the data transmission and control of the detector, two uplinks of 10\,Gigabit are foreseen, providing sufficient margin for the future detector readout. Online data processing for noise suppression will be performed before data storage. The algorithm to perform the noise filtering can be based on the coincidence of hits in time and planes.  
\subsubsection{Equipment control and monitoring}
A small amount of monitoring and remote controlled equipment will be located in the detector rack. A common cooling unit delivering water cooling for the SciFi and a air cooling for the Emulsion detector is remote controlled via Ethernet. All power supplies for the Timing detector and SciFi providing bias voltage for SiPMs and low voltage for the front end electronics is controlled and monitored via Ethernet as well as all additional temperature or radiation monitoring systems. It is not foreseen to have any local CPU located in the rack. 

\section{Detector integration in the TI18 cavern}
\label{subsec:integration}

Particular effort has been paid on the studies for the integration of the needed services, to guarantee a cost efficient design of the SND detector and to evaluate the feasibility of its installation inside the TI18 cavern.
A preliminary study of the infrastructure requirements was performed and it was followed by a first iteration with the concerned CERN groups, in order to define the best installation strategy and the  project requirements. 
A first proposal has been presented at the Int\'egration Cellule LHC meeting, which is in charge of the integration studies of the LHC machine, on February 12th 2020: no showstoppers and integration issues were identified. 

\begin{figure}[ht]
  \centering
  \includegraphics[width=\linewidth]{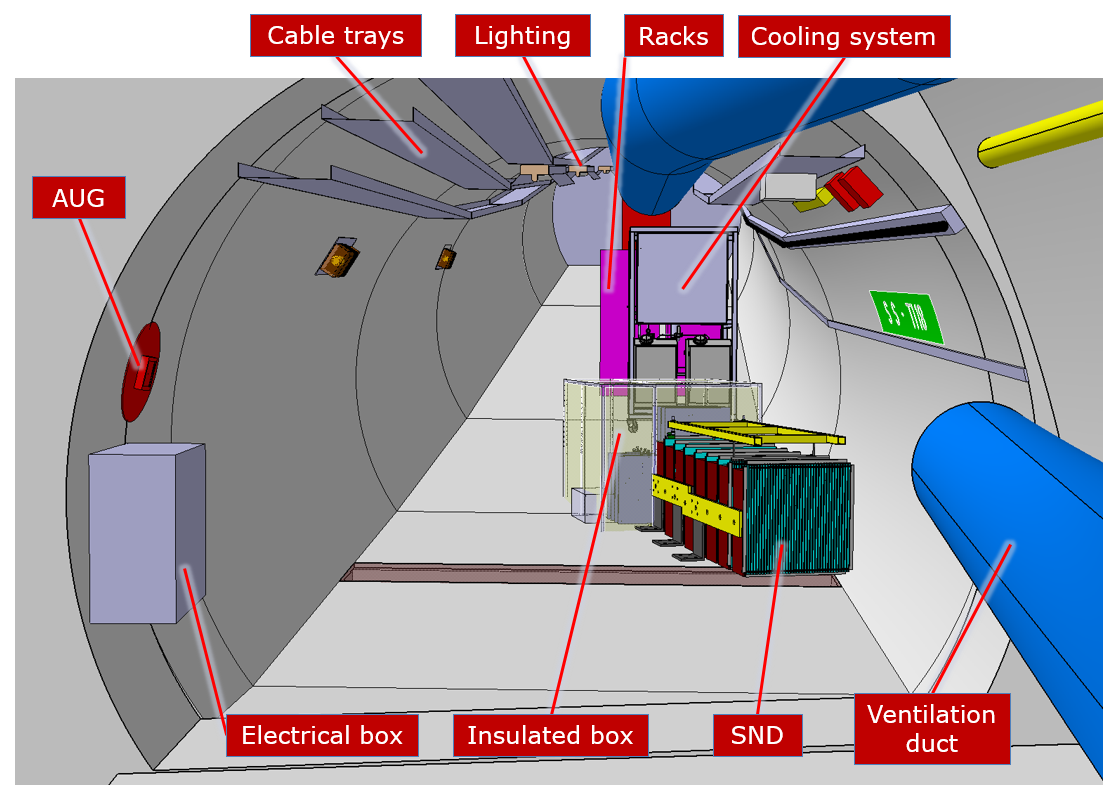}
  \caption{SND integration inside TI18}
  \label{fig:integration}
 \end{figure}

Preliminary survey, electrical engineering, cooling and ventilation, transport, safety and radiation protection specifications were used to achieve the first integration of the SND detector as shown in Figure~\ref{fig:integration}.

\begin{figure}[ht]
  \centering
  \includegraphics[width=\linewidth]{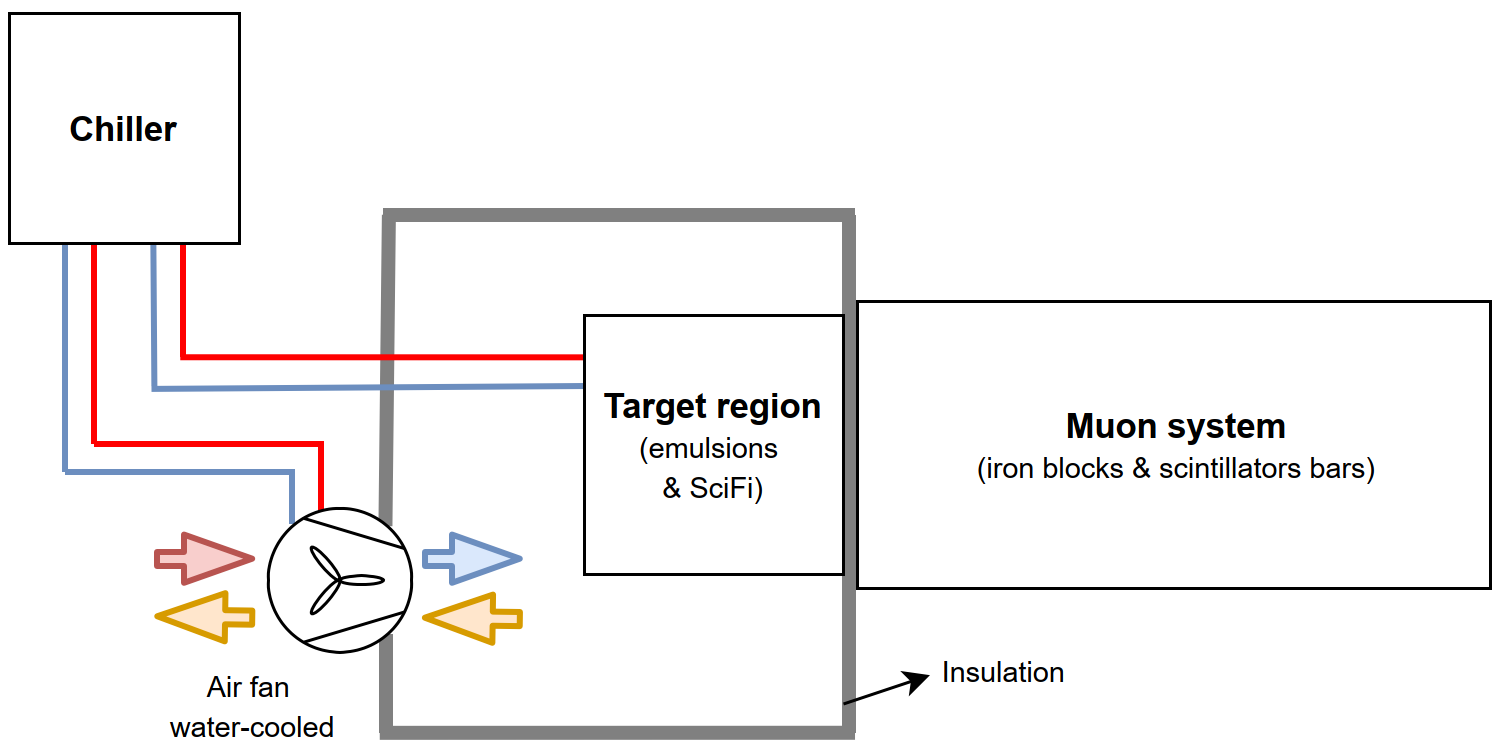}
  \caption{Proposed cooling scheme for the SND detector}
  \label{fig:cooling}
 \end{figure}

As described in Section \ref{subsec:ECC}, the temperature of the emulsion target must be below \SI{15}{\celsius} in order to avoid possible fading effects. In addition, the SciFi needs \textbf{water-cooling} for its electronics. Therefore, the proposal includes two chillers located next to the detector and an insulated box equipped with an air fan. The chillers will provide cooled water to the electronics and to the air fan. The later will cool down the refrigerated volume taking the air from the tunnel. Figure~\ref{fig:cooling} shows the conceptual  scheme of the cooling system. A mineral wool layer or a fire barrier duct wrap can be used as insulator and aluminium profiles to build the structure of the insulated box. The structure should include another layer of boron carbide to keep the detector protected from thermal neutrons. The expected water-cooling power consumption is~\SI{2.5}{\kilo\watt}.

\begin{figure}[ht]
  \centering
  \includegraphics[width=\linewidth]{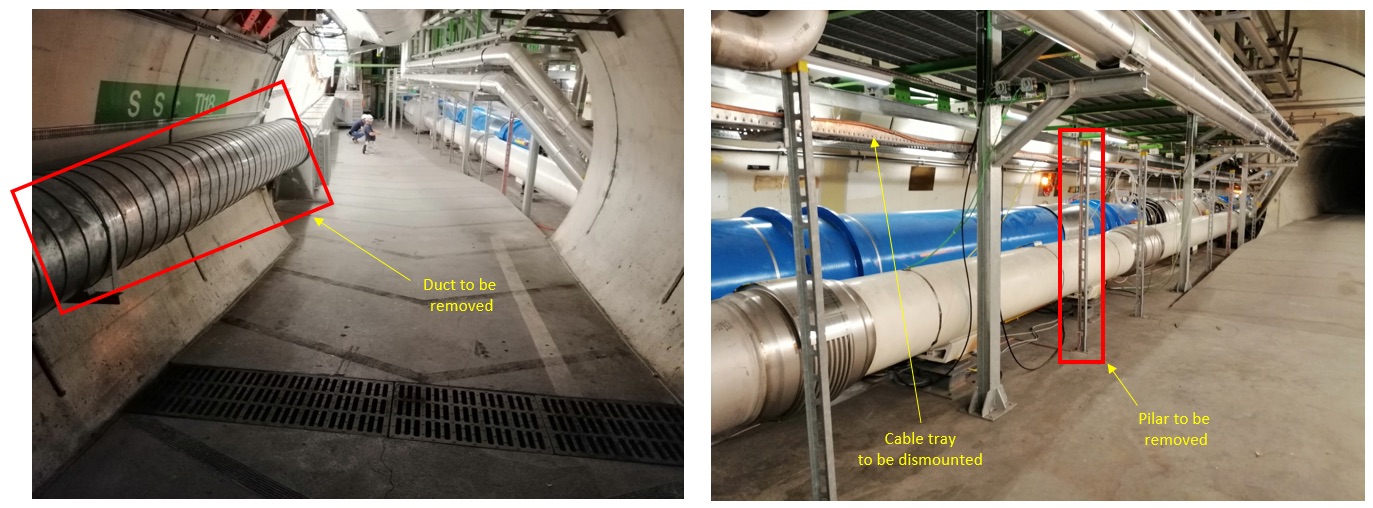}
  \caption{Activities to be carried out for the ventilation duct removal}
  \label{fig:venductremoval}
 \end{figure}

It has been identified that a \textbf{ventilation} duct located inside the TI18 cavern has to be removed to avoid interference with the detector. A visit with EN-CV, RP, transport and the coordination teams was organised on December 2019 where it was concluded that the duct can be easily removed, with slight modifications on the existing infrastructure, and transported to the surface (see Figure~\ref{fig:venductremoval}). Samples will be taken and analysed by HSE-RP to check the activation of the ventilation duct prior its removal. Nevertheless, from the experience of similar works conducted in TI12, the activation of such components is expected to be below the clearance limits (LL).

The sub-components and services required to run the detector can be transported by a trolley from the surface level to the LHC tunnel through the PM15 elevator and, then, transported along the LSS1R to UJ18. Inside UJ18 the small parts can be passed below the LHC machine with a low trolley, where there is a free passage about \SI{80}{cm} wide and \SI{34}{cm} high, and finally, install them in TI18. However, there are a few components  which have to be passed  above the LHC owing to their dimensions. Therefore, a hoist  with about \SI{250}{\kilogram} capacity is required. 
A protection structure (similar to FASER-UJ12 protection~\cite{LHCQRMPI0001}) has to be installed for the protection of the LHC machine and cryogenic line. 
For the final installation of the heaviest parts, a specific handling equipment is required to climb the slope of the TI18 cavern. After discussion and a site visit with the \textbf{transport} experts, no showstoppers have been identified. Table~\ref{tab:transport} includes the weight and dimensions of the most restrictive parts to be handled.  

\begin{table} [htpb]
  \centering
  \caption{The most restrictive sub-components to be transported}
  \label{tab:transport}
\begin{tabular}{lccc}
\hline
\multicolumn{1}{c}{\textbf{Part}} & \textbf{Dimensions (cm)} & \textbf{Unit weight (kg)} & \textbf{Number of units} \\ \hline
\textit{Veto plane} & 100 x 100 x 25 & 20 & 1 \\
\textit{Muon detector} & \multicolumn{1}{l}{} & \multicolumn{1}{l}{} & \multicolumn{1}{l}{} \\
\multicolumn{1}{c}{\begin{tabular}[c]{@{}c@{}}SAMPIC boards\\    SAMPIC Power Supply\\    HV+LV supplies, crate\\    Veto plane\\    Muon timing plane\\    Muon x,y plane\end{tabular}} & \begin{tabular}[c]{@{}c@{}}50 x 50 x 30\\ 50 x 30 x 30\\ 50 x 50 x 30\\ 60 x 60 x 20\\ 80 x 80 x 20\\ 80 x 80 x 20\end{tabular} & \begin{tabular}[c]{@{}c@{}}10\\ 8\\ 20\\ 4\\ 6\\ 8\end{tabular} & \begin{tabular}[c]{@{}c@{}}6\\ 1\\ 1\\ 1\\ 5\\ 6\end{tabular} \\
\textit{SciFi} & \multicolumn{1}{l}{} & \multicolumn{1}{l}{} & \multicolumn{1}{l}{} \\
\multicolumn{1}{c}{\begin{tabular}[c]{@{}c@{}}Modules\\    Power supplies, crates, switch\end{tabular}} & \begin{tabular}[c]{@{}c@{}}100 x 100 x 25\\ 50 x 50 x 30\end{tabular} & \begin{tabular}[c]{@{}c@{}}15\\ 20\end{tabular} & \begin{tabular}[c]{@{}c@{}}4\\ 6\end{tabular} \\
\textit{Iron block} & 80 x 20 x 20 & 230* & 32 \\
\textit{Rack} & 50 x 50 x 170 & 25 & 2 \\
\textit{Chiller} & 100 x 60 x 40 & 80 & 2 \\
\textit{Emulsion brick} & 39.7 x 42.6 x 8.5 & 9 & 4 \\ \hline
\multicolumn{4}{l}{*It can be divided in smaller pieces to reduce weight and can pass below the LHC machine}
\end{tabular}
\end{table}

The conventional~\textbf{safety} aspects have been discussed with HSE-OHS group and no showstoppers have been identified. A PESS-correspondent has been assigned to follow the project and evaluate, together with the HSE-OHS specialists, the safety aspects in detail within the Coordination of Project Experiment Safety Support activities. Regarding the personnel space route, there is already installed a bridge above the magnet MBB13.R1 to pass over the LHC machine in case of an emergency. The Launch Safety Discussion document is under preparation in order to proceed with the release of the safety project requirements report. 

A detailed~\textbf{radiation protection} study has been launched by the HSE-RP to evaluate the radiation levels during TS and the activation of the SND detector’s components within the TI18 cavern, considering run 3 conditions. HSE-RP will provide a detailed radiological evaluation of the SND detector operation and maintenance in the coming months. The access of personnel over 1 or 2 days during the technical stop to replace the emulsion bricks seems feasible due to the relatively low activation risk expected in TI18 during Run 3. The exposure time of the bricks to the cosmic rays has to be minimised, therefore, an effective plan among the transport, RP and detector teams is to be developed.  

Radiation levels in TI18 are dominated by beam-gas interaction from the incoming beam, with the luminosity-driven losses in cells 11 and 13 of the dispersion suppressor for the outgoing beam being negligible. The measured radiation values during 2018~\cite{R2E}, scaled to a nominal Run 3 operation year, yield expected annual levels of 2e7 and 1e8 cm$^{-2}$yr$^{-1}$ HEHs and thermal neutrons, respectively. Such levels are almost an order of magnitude larger than the limits which EN-STI-BMI considers to declare an area as radiation safe for pure commercial electronics systems. However, it is to be noted that the limits consider distributed systems, whereas for a single unit system, they can be regarded as clearly pessimistic; and that the measured levels correspond to the intersection between TI18 and the LHC tunnel (UJ18) and equipment would clearly benefit from being positioned some meters inside TI18. Therefore, the prospects are that commercial electronics in TI18 are expected to operate in radiation safe conditions, however, a more detailed study of the environment (i.e.~2D FLUKA maps with R2E quantities) and electronics type description would need to be carried out further ahead in the project for a final confirmation.

There is a clear line of sight to link the detector position with the LHC machine coordinate system. In order to do so, three \textbf{survey} fiducials per element to be measured or aligned are to be included in the detector set-up. They have to be as far away as possible from each other to ensure good quality measurements and they must be visible from a single fixed position. Every component that has to be aligned should be supported isostatically on 3 points. The first approach indicates that an absolute alignment and measurement accuracy of \SI{1}{\milli\meter} is sufficient for the SND detector purposes. The alignment and measurement accuracy among sub-components (i.e. SciFi plane, emulsion plane etc.) has been set to \SI{0.2}{\milli\meter}.  

\begin{figure}[htp]
  \centering
  \includegraphics[width=\linewidth]{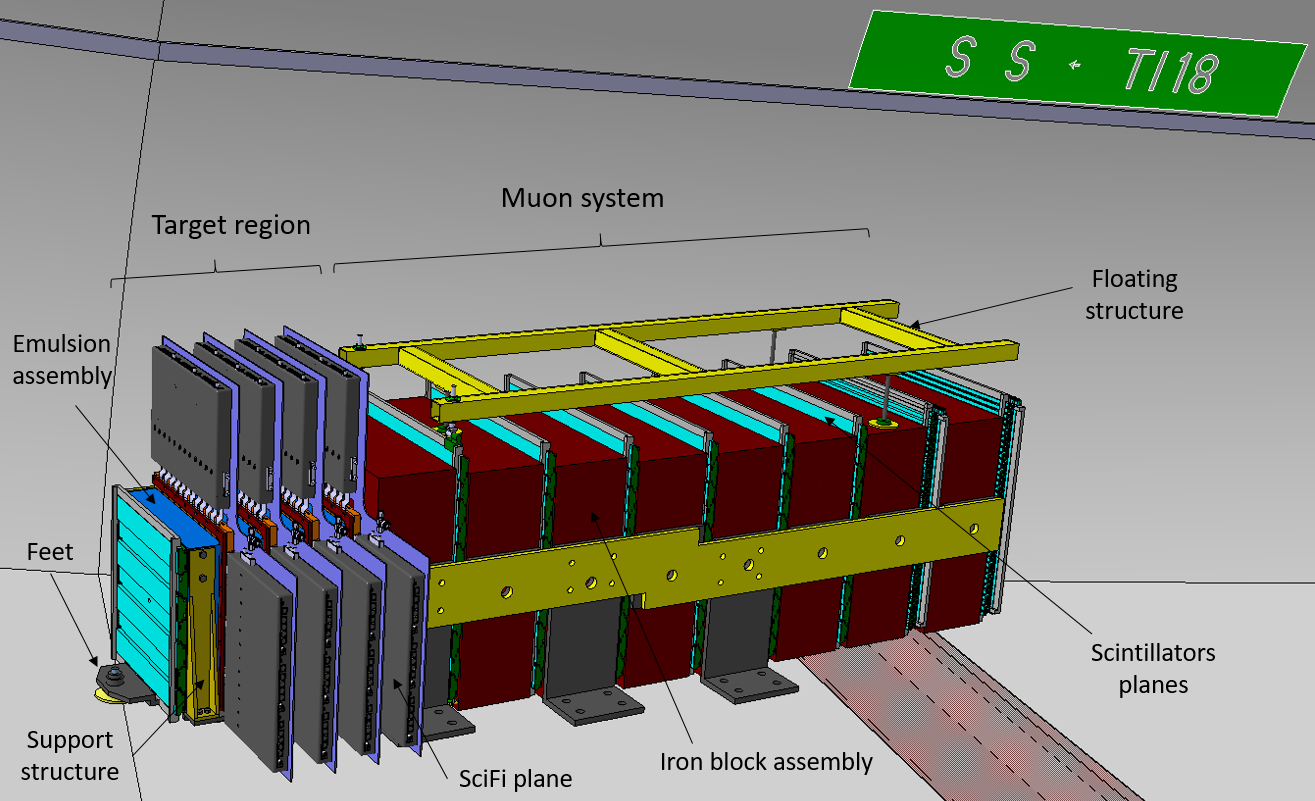}
  \caption{Detector layout pointing out the main concepts of the alignment system}
  \label{fig:su}
 \end{figure}

A first conceptual design (see Figure~\ref{fig:su}) of the alignment strategy divides the detector in two independent assemblies:

\begin{itemize}
    \item The target region is supported by a single support structure and aligned by three feet (the SPS magnet feet design \cite{surveyfeet} has been taken as reference). A kinematic system mechanically fixed by three points (a slot, a cone and a flat) to the support structure has to be designed for the emulsion and SciFi planes. Each wall will have a surface on the corner to place the survey fiducials. An emulsion wall is made of 12 bricks to be aligned mechanically by their own assembly frame. They can be fiducialized by a Romer arm on the surface if clear references (fiducial and/or surfaces) are included on the bricks.  
    \item The muon system is supported by the iron blocks. A plate with three screws will be attached to the floor and aligned. By filling the space between the plate and the floor with concrete, the plate will be rigidly fixed to the floor. This operation will be done for the assembly of each iron block. After that, the iron blocks are piled up and positioned by space holders to keep the required space for the insertion of scintillators bar planes. On top of the assemblies and supported by three points fixed to two different iron blocks, there is a floating structure which will support the scintillator bar planes. Each scintillator bar plane is supported by three points (two on top and one on the side) and will need an available surface for the installation of fiducials. 
\end{itemize}

The alignment system of the detector is being discussed with the alignment experts to find the best design complying with the specifications and guarantee a robust supporting system. The local gravity direction has been provided by the survey group for design purposes.

The preliminary requirements and \textbf{electrical services} have been identified in order to comply with the detector specifications. The total electrical power of \SI{10}{\kilo\watt} will be enough to supply the chillers, lighting equipment, SciFi, timing detector and racks. The electrical services addressed include: an electrical distribution box located at the entrance of the TI18 cavern, two AUG required for safety aspects, permanent lighting (about 3 normal lamps plus two anti-panic lighting) and x5 sockets of \SI{220}{\volt}.

\begin{figure}[ht]
  \centering
  \includegraphics[width=\linewidth]{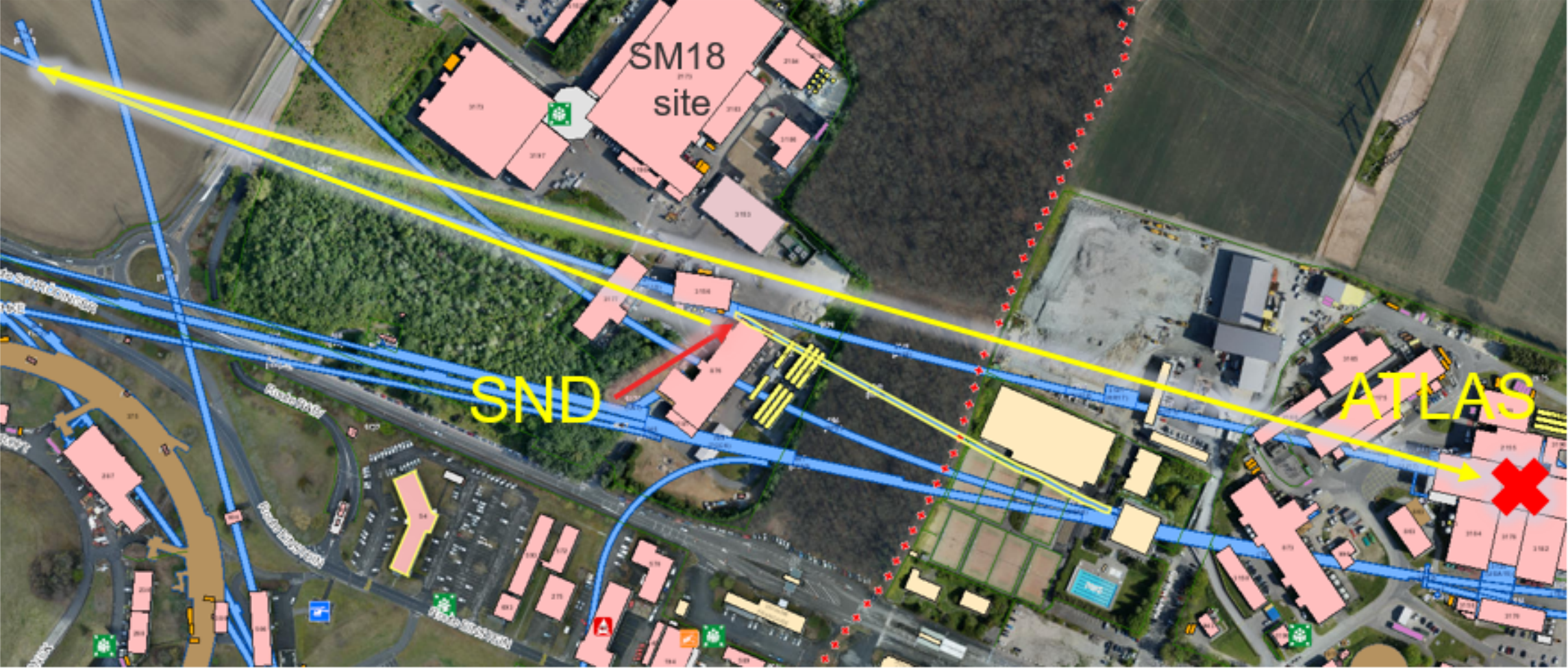}
  \caption{Preliminary optical fibres layout}
  \label{fig:optfib}
 \end{figure}

In relation to the signals, readout and networking infrastructure, the following items have been identified:

\begin{itemize}
    \item Two 19' racks for the fibre optics connection modules, NIM crates and power supply for the detector components.
    \item A rack on surface to host the server for data acquisition. 
    \item A set of 12 fibre optics for the detector, Ethernet and cooling control system communication. The preliminary optical fibres layout is shown in Figure~\ref{fig:optfib}. It was considered preferable to avoid passing them along the Long Straight Section to avoid their de-installation and re-installation due to the HL-LHC upgrade during LS. 
\end{itemize}

Very productive and effective discussions with EN-SMM, HSE-OHS, HSE-RP, EN-HE, EN-EL-EIC, EN-EL-FC, IT-CS-DO, EN-CV, IT/CS/CS, EN-STI-BMI concluded that the preliminary SND integration and installation of the required infrastructure are feasible.
\section{Emulsion data analysis}
\label{sec:charm}

\subsection{Emulsion scanning and analysis}
Emulsion films consist of two 70$\mu$m-thick layers of nuclear emulsion, separated by a 175~$\mu$m-thick plastic base.
The transverse size is $12.5 \times 10$~cm$^2$, like for the passive plates.

The track left by a charged particle on an emulsion layer is recorded by a series of sensitisied AgBr crystals, growing up to 0.6 $\mu$m diameter during the development process. A new generation automated optical microscope analyses the whole thickness of the emulsion, acquiring various topographic images at equally spaced depths. The acquired images are digitized, then an image processor recognizes the grains as  \textit{clusters}, i.e.~groups of pixels of given size and shape. Thus, the track in the emulsion layer (usually referred to as \textit{microtrack}) is obtained connecting clusters belonging to different levels. Since an emulsion film is formed by two emulsion layers, the connection of the two microtracks through the plastic base provides a reconstruction of the particle's trajectory in the emulsion film, called \textit{base-track}.  The full-volume wide reconstruction of particle tracks requires connecting base-tracks in consecutive films. In order to define a global reference system, a set of affine transformations has to be computed to account for the different reference frames used for data taken in different films.\\
The analysis of emulsion films is performed using new generation optical microscopes. For the OPERA experiment two different scanning systems were developed: one in Japan, the Super Ultra Track Selector (S-UTS) \cite{Morishima:2010zz}, and one by a collaboration of the different European laboratories, the European Scanning System (ESS) \cite{Armenise:2005yh, Arrabito:2006rv, Arrabito:2007td, DeSerio:2005yd}. The ESS is a microscope equipped with a computer-controlled monitored stage, movable along both X and Y axes and in the Z direction, a dedicate optical system and a CMOS Mega-pixel camera mounted on top of the optical tube. For each field of view, it executes the following steps: local tomography, cluster recognition, grain selection, three-dimensional reconstruction of aligned cluster grains, parameter extraction for each grain sequence. The ESS allows the scanning of the whole emulsion  surface with a maximum speed of 20 cm$^2$/h.\\
An upgrade of the ESS system was  performed by the Naples emulsion scanning group \cite{Alexandrov:2016tyi}. The use of a faster camera with smaller sensor pixels and a higher number of pixels combined with a lower magnification objective lens, together with a new software LASSO \cite{Alexandrov:2016tyi,Alexandrov:2015kzs} has allowed to increase the scanning speed to 180 cm$^2$/h \cite{Alexandrov:2017qpw}, more than a factor ten larger than before. The lens of the microscope guarantees a submicron resolution and, having a working distance in Z longer than 300 $\mu$m, to scan both sides of the emulsion film. To make the optical path homogeneous in the film, an immersion lens in an oil with the same refraction index of the emulsion is used. A single field of view is 800$\times$600 $\mu$m$^2$; larger areas are scanned by repeating the data acquisition on a grid of adjacent fields of view. The images grabbed by the digital camera are sent to a vision processing board in the control workstation to suppress noise.

\subsection{The SHiP-charm measurement}

The SHiP Collaboration proposed the SHiP-charm project~\cite{Akmete:2286844}, aiming at measuring the associated charm production induced by 400 GeV/c SPS protons. This proposal includes a study of the cascade effect~\cite{CASCADE} to be carried out using the ECC technique, i.e.~slabs consisting of a replica of the SHiP experiment target~\cite{Anelli:2015pba} interleaved with emulsion films. The detector is a hybrid system, combining the emulsion technique with electronically read out detectors and a spectrometer magnet to provide the charge and momentum measurement of charmed hadron decay daughters and the muon identification.

An optimisation run was performed at the  H4 beam-line of the CERN SPS in July 2018 with an integrated number of protons on target of about $1.5 \times 10^6$. Lead was used as passive material for the ECC. Six different target configurations were used in order to measure the charm production along a depth corresponding to $\sim$2 interaction lengths: in this depth 
about 80\% (50\%) of the charm production from primary (secondary) interactions is expected. 

\begin{figure}[h!]
\centering
\includegraphics[width=0.9\linewidth]{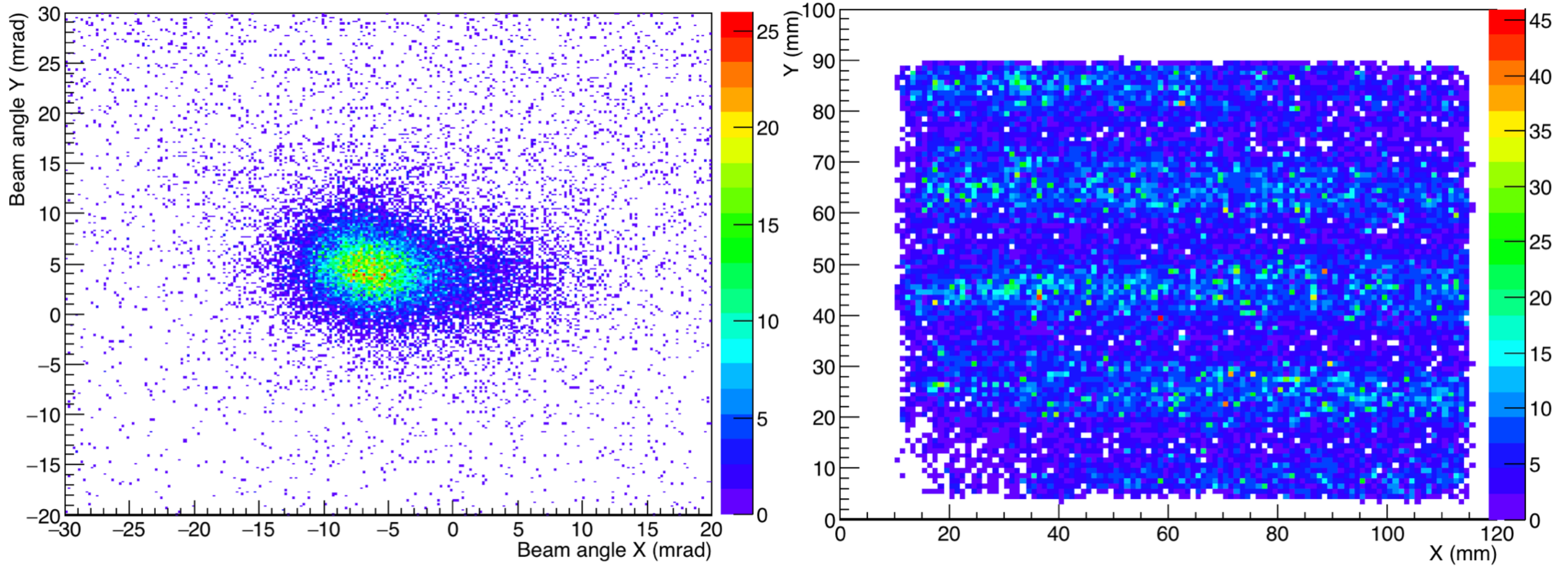}
\caption{Left: angular dispersion of the proton beam as reconstructed in one of the exposed ECC target units. Right: position distribution of the incoming protons on the emulsion surface.}
\label{fig:charm_beam}
\end{figure}

Each ECC was mounted on a moving table in order to change the position of the target with respect to the proton beam and irradiate the whole surface of the detector. Figure ~\ref{fig:charm_beam} reports the characterisation of the proton beam in one of the ECC targets both in terms of angle (left) and position (right). The pattern observed in the position distribution reproduces the movement of the target with respect to the proton beam.

\begin{figure}[h!]
\centering
\includegraphics[width=0.9\linewidth]{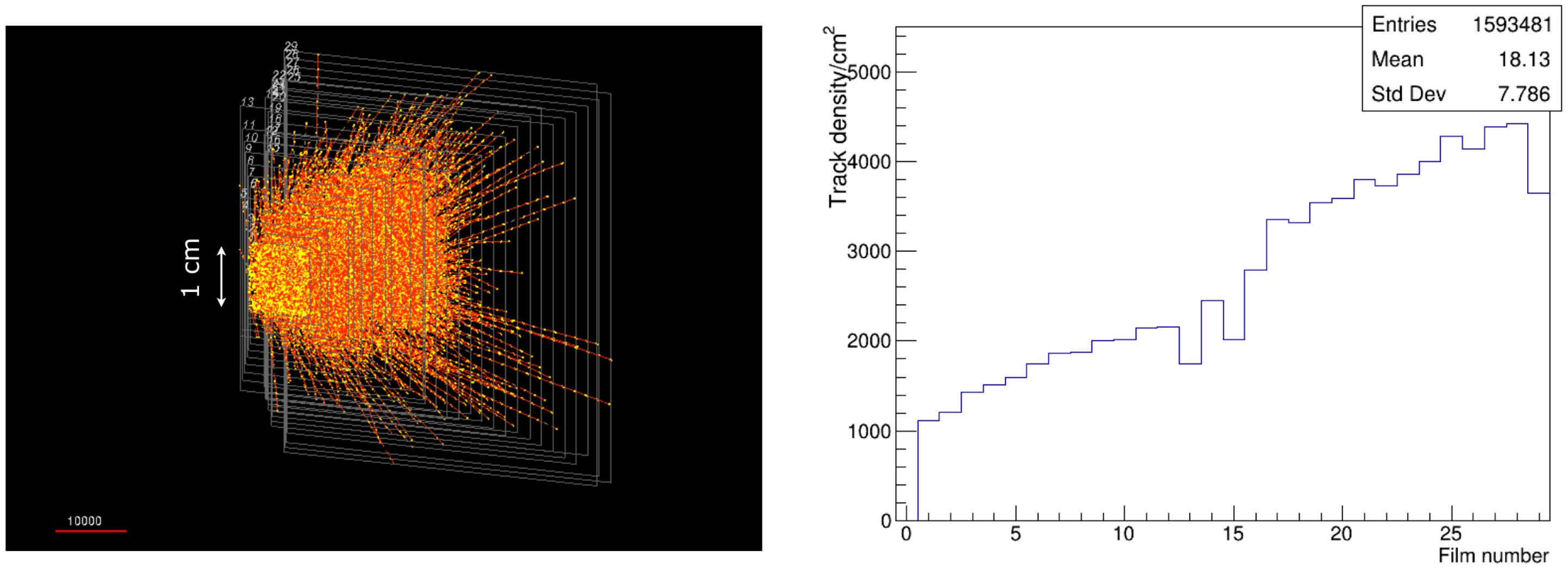}
\caption{Left: Tracks reconstructed in a 1$\times$1 cm$^2$. Right: track density in one of the exposed ECC target units.}
\label{fig:charm_density}
\end{figure}

The number of protons impinging on ECC target units vary from 10$^2$/cm$^2$ to 10$^3$/cm$^2$ according to the configuration of the exposure. The data analysis shows that the track density increases with the depth in the module due to the proton interactions, hadronic re-interactions and electromagnetic showers, as shown in Fig.~\ref{fig:charm_density}. The density of segments reconstructed in a single emulsion film extends up to 5$\times$10$^4$/cm$^2$.  
The challenge of the SHiP-charm measurement is two-fold: reconstruct tracks and interaction vertices in a high density environment, and search for rare decays of charmed hadrons. The first challenge is also shared by the SND@LHC measurement where we expect to get an integrated density of track segments up to 10$^5$/cm$^2$ for an exposure of about 25~fb$^{-1}$. 

About 10$^4$ proton interaction vertices are expected in a single ECC target unit ($\sim$10$^3$/cm$^3$). 400 GeV/c proton interactions produce on average 15 charged particles and 10 photons, having energies ranging from a few to tens of GeV. This results in a large number of secondary hadronic re-interactions and electromagnetic showers, that increases the number of reconstructed vertices up to 5$\times$10$^5$.
To set the scale,  the unitary target of the OPERA experiment contained in the same volume a single neutrino interaction vertex.

\begin{figure}[h!]
\centering
\includegraphics[width=0.9\linewidth]{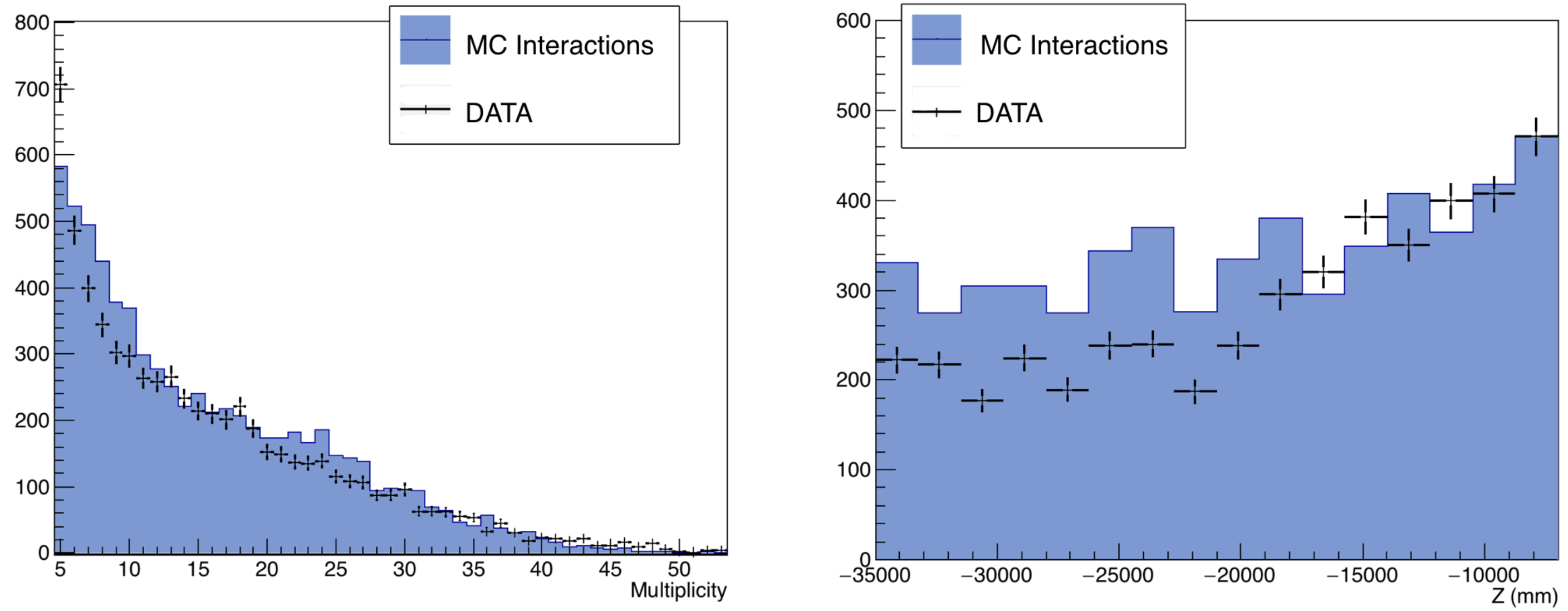}
\caption{Charged track multiplicity and position along the beam direction for interaction vertices reconstructed in one of the ECC target units. Data points are in good agreement with Monte Carlo expectations.}
\label{fig:charm_primaryvertex}
\end{figure}

The analysis of the SHiP-charm emulsion data therefore required the development of dedicated software and analysis tools to extract the signal from an unprecedented background rate. The results obtained for the reconstruction of proton interaction vertices in one of the ECC target units are reported in Fig.~\ref{fig:charm_primaryvertex}. A good agreement between data and Monte Carlo expectations is found, both in normalisation and in shape, for the number of charged tracks defining the interaction vertex and the position of the vertex along the beam axis. These results prove the capability to reconstruct interaction vertices in a harsh environment. 

\begin{figure}[h!]
\centering
\includegraphics[width=0.9\linewidth]{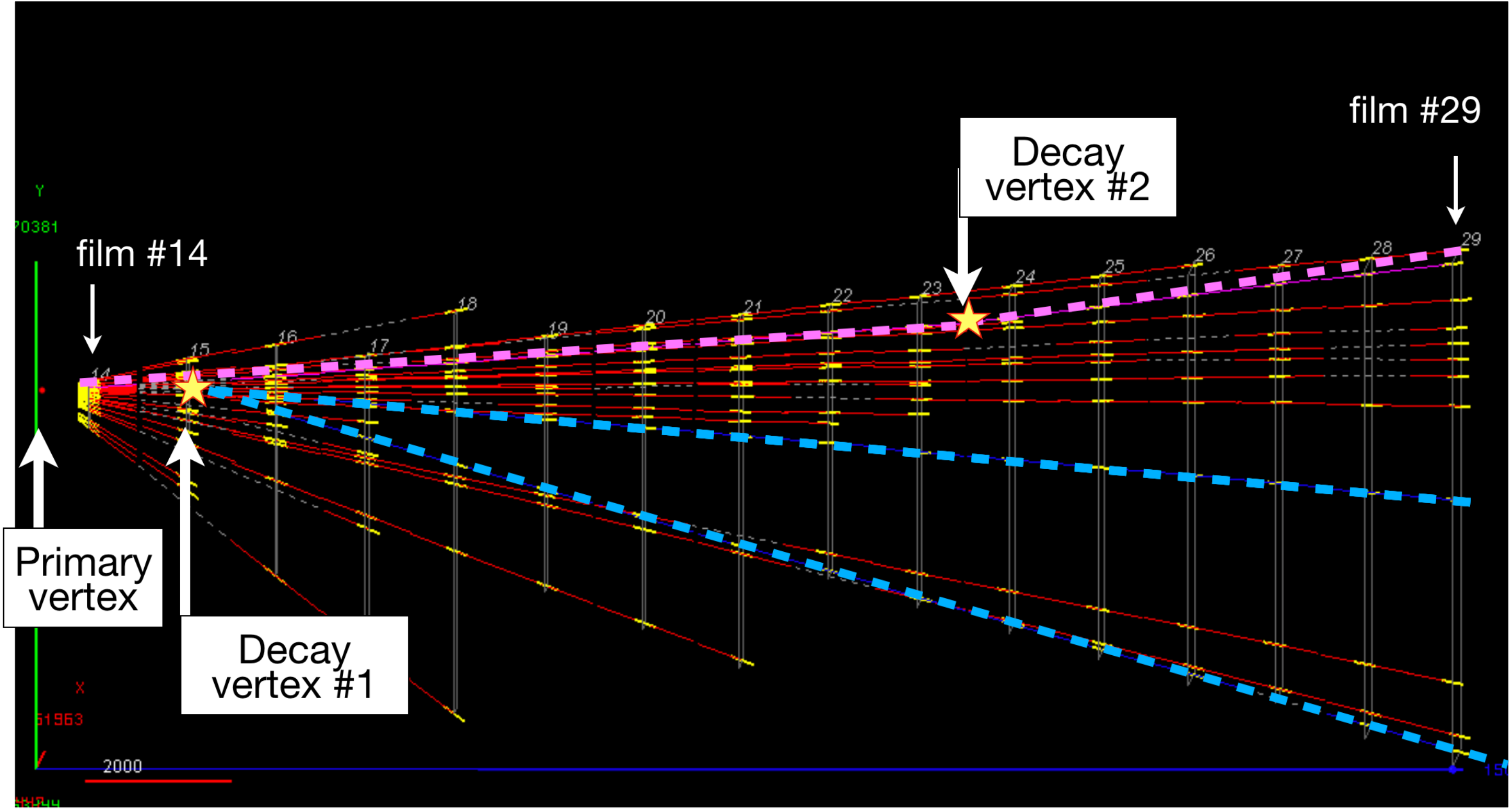}
\caption{One of the double-charm candidates reconstructed in the ECC.}
\label{fig:charm_candidate}
\end{figure}

The search for charmed hadrons is based on the observation of a decay topology in a $\sim$55 mm$^3$ volume downstream of each interaction vertex and relies on the measurement of topological and kinematic variables at the secondary vertex. The required background suppression is of the order of 10$^4$ and specific analyses are being developed to enhance the signal-to-noise ratio. Figure~\ref{fig:charm_candidate} shows an example of a double-charm candidate event that has three vertices: the most upstream is the proton interaction vertex. There are two additional vertices: one shows a two-prong topology without any charged parent particle while the other one shows a kink topology. 

Although signal and backgrounds expected at SND@LHC are different from those from the SHiP-charm measurement, the two projects share the same challenging task of identifying interactions in an unprecedented flux of charged particles. The analysis tools recently developed by the SHiP collaboration for the SHiP-charm project can be therefore applied for the analysis of emulsion data in SND@LHC.

\section{Physics performance}

\subsection{Simulation Software}

The simulation of the SND@LHC detector is handled by the FairShip software suite, developed within the SHiP collaboration, which is based on the FairRoot software framework~\cite{FAIRROOT}.

Neutrinos production in proton-proton collisions at the LHC is simulated
with DPMJET3 (Dual Parton Model, including charm)~\cite{Roesler_2001}  embedded in FLUKA~\cite{fluka, fluka2}, and particle propagation towards the detector is done through the FLUKA model of the LHC accelerator in order to simulate also neutrinos from further decay of collision and reinteraction products.
GENIE~\cite{cite:GENIE} is then used to simulate neutrino interactions with the detector material.
The output of GENIE is given to {\sc Geant4}~\cite{Geant4} for the particle propagation in the detector.

\subsection{Neutrino physics}
\label{subsec:neutrino}

\begin{figure}
 \centering
  \subfloat[]{\includegraphics[width = 0.45\textwidth]{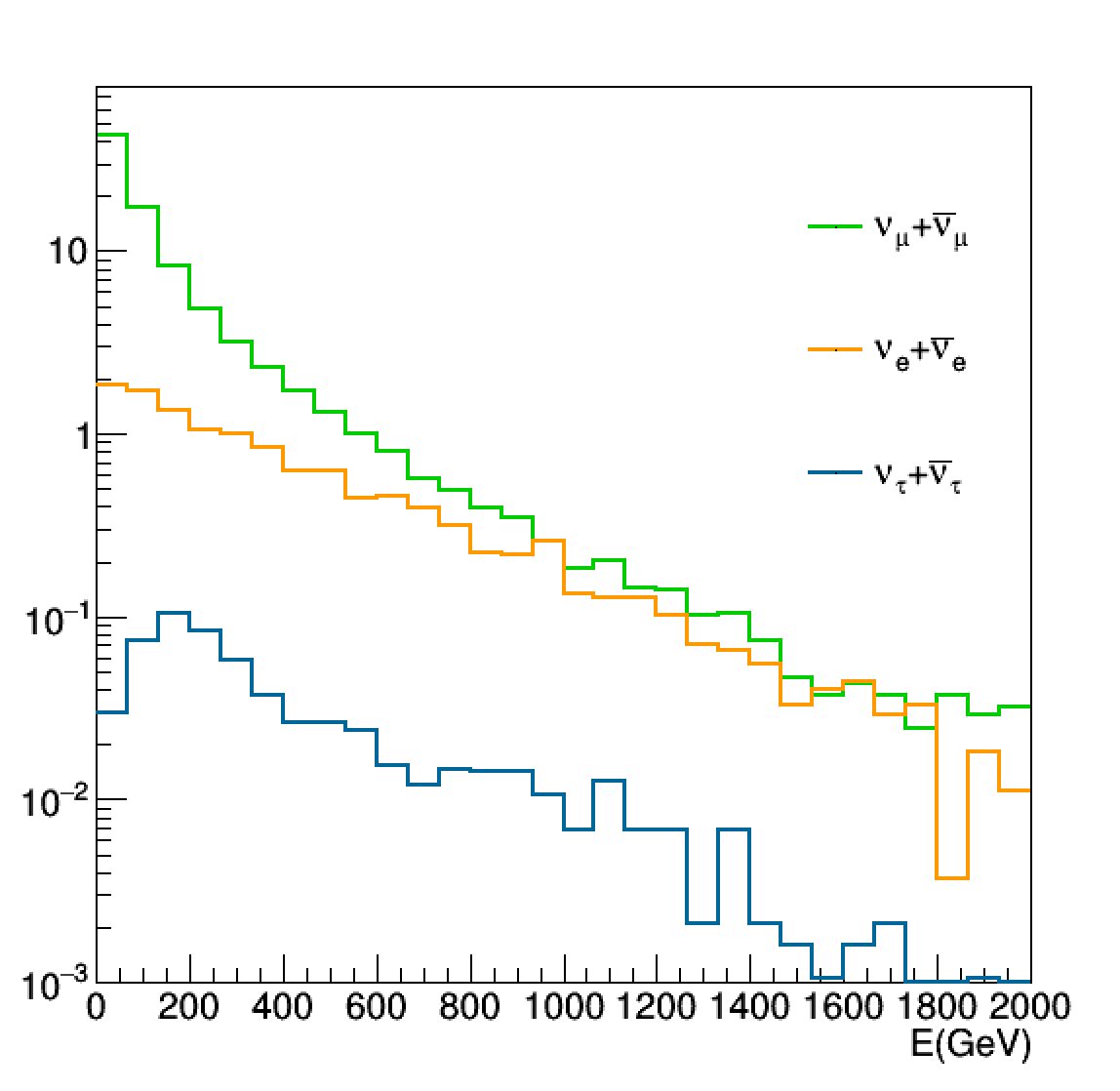}\label{sf:spectrumTarget}}
\hfill
\subfloat[]{\includegraphics[width = 0.45\textwidth]{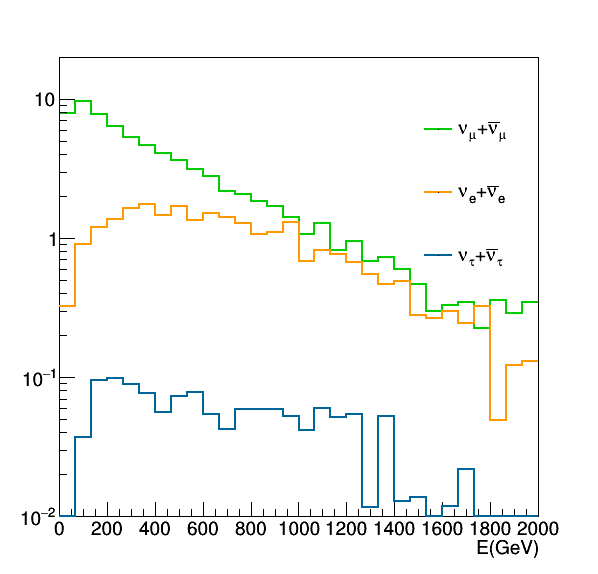}\label{sf:spectrumCC}}
 \caption{Energy spectra of the three neutrino flavours at the target region (a) and interacting in the target region (b). The total number of neutrinos is normalised to 100.}
 \label{fig:spectra}
 \end{figure}

The neutrino detector covers a pseudo-rapidity range between $7.2<\eta<8.7$.
The spectra and the neutrino yield at the neutrino detector for the three different neutrino flavours are reported in Fig.~\ref{sf:spectrumTarget} and in the left column of 
Table~\ref{tab:NuFlux}, respectively. The muon rate is expected to be 100 Hz over the whole detector surface.

Assuming a luminosity of 25 fb$^{-1}$ for the initial configuration, the integrated  neutrino flux for the different neutrino flavours at the target region is reported in the left column of Table~\ref{tab:NuFlux} and the corresponding energy spectra are shown in Fig.~\ref{sf:spectrumTarget}.
The expected number of charged-current neutrino interactions occurring in the detector is reported in Table~\ref{tab:NuFlux} for the initial configuration, while the energy spectra are shown in Fig.~\ref{sf:spectrumCC}.
The last column of Table~\ref{tab:NuFlux} reports the expected yield of neutrino charged-current interaction in the updated  configuration of the detector, assuming an integrated luminosity of 150 fb$^{-1}$.

\begin{table}[h!]
 \caption{(Third column) Integrated neutrino flux for 25 fb$^{-1}$ for the  different neutrino flavours at the target region. (Fifth column) Expected number of CC interactions for the different neutrino flavours for 25 fb$^{-1}$. (Sixth Column) Expected number of CC interactions for the different neutrino flavours for 150 fb$^{-1}$ in the updated detector configuration.}
  \label{tab:NuFlux}
\centering
\begin{tabular}{c  c c  c c c}
\toprule
Neutrino &    $\langle\text{E}\rangle   $  & Neutrino &   $\langle\text{E}\rangle$   & CC & CC \\
 flavour &   GeV    & Flux &    GeV    & Interactions & Interactions\\
 & (incident) &  & (interacting) & Initial config & Updated config\\
\midrule
  $\nu_\mu$ & 150 & $4.6 \times 10^{11}$ &  460  & 62 &975\\
   $\nu_e$     & 390 & $5.9 \times 10^{10}$ & 710 & 21 &332\\
    $\nu_\tau$ & 420 & $3.0 \times 10^{9}$ & 720 & 1 & 18\\
  $\bar{\nu}_\mu$  & 150 & $4.0 \times 10^{11}$   & 480 & 27 & 429\\
   $\bar{\nu}_e$ & 390 &$6.2  \times 10^{10}$  & 740  & 11 & 174\\
    $\bar{\nu}_\tau$   & 360 & $2.9 \times 10^{9}$  & 720 & 0 & 7\\
\midrule
 TOT & & $9.87 \times 10^{11}$ & & 122 & 1935\\
\bottomrule
 \end{tabular}
 \end{table}
 
\subsubsection{Neutrino detection}

For the data analysis a fiducial volume inside the brick is selected, excluding the regions within 5 mm from the downstream edge and 1 mm from the lateral side.
This results in a geometrical efficiency $\epsilon_{geom}$ of approximately 89.3\%.
The location of a neutrino interaction requires the presence of least one track with a momentum lager than 1 GeV/c attached to the primary vertex identified by two visible tracks in emulsion. 
The visibility criteria are listed in Tab.~\ref{tab:VisCut}. Monte Carlo simulation studies, fully validated with data from the SHiP-charm measurement, have shown that the location efficiency for the different neutrino flavours is above 90$\%$.

\begin{table}
\centering
\begin{tabular}{c  c }
\toprule
Particle &    Visibility cut p (MeV/c)\\
 \midrule
  charged     & $>100$\\
  $p$ & $>170$\\
  $\pi^0$ & $>400$\\
  $\gamma$ & $>200$\\
 \bottomrule
 \end{tabular}
 \caption{Visibility cuts on the momentum of primary tracks at the neutrino vertex.}
  \label{tab:VisCut}
 \end{table}

\paragraph{Muon Identification}

\begin{figure}
    \centering
    \includegraphics[width = 0.75\textwidth]{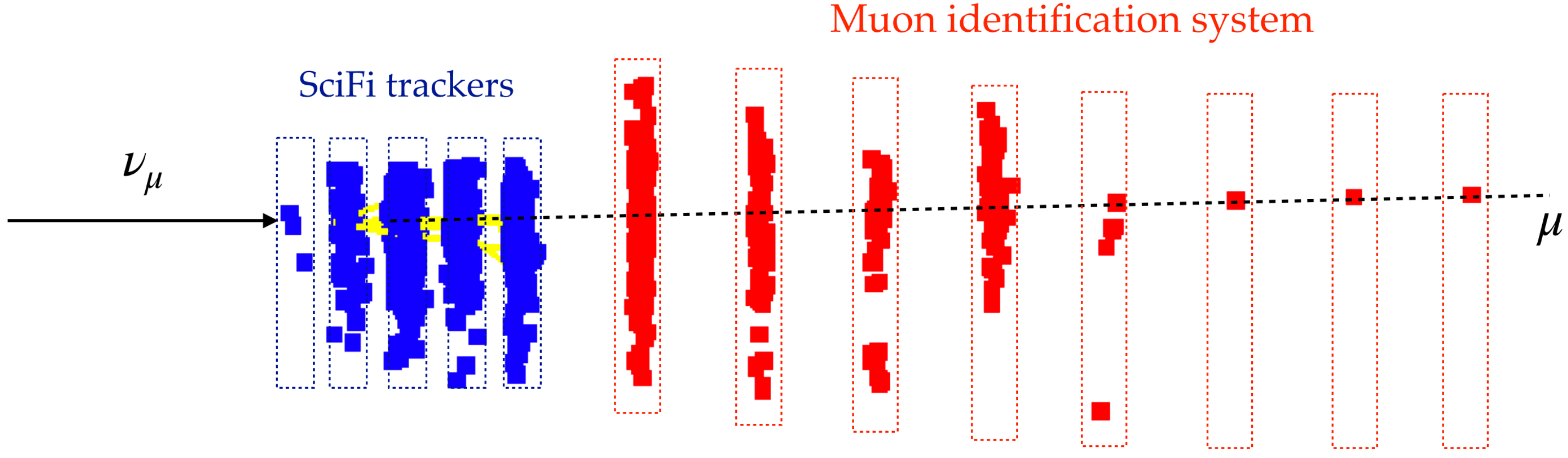}
    \caption{Simulated muon neutrino interaction. The muon exhibits isolated hits in the three most downstream planes of the muon identification system.}
    \label{fig:eventDisplay}
\end{figure}

Muon identification  is relevant for both identifying $\nu_\mu$ and   $\nu_\tau$ interactions.  As a matter of fact, charmed hadrons produced in $\nu_\mu$~CC interactions constitute a background for the $\nu_\tau$ search if the primary muon is not identified.
A display of a simulated $\nu_\mu$ CC interaction can be found in Fig.~\ref{fig:eventDisplay}.

MonteCarlo studies have shown that $\approx 98.4\%$ of the muons produced in $\nu_\mu$ CC interactions enter the muon system.
Out of them, $\approx 91.5\%$ leave a hit in the last three planes of the muon identification system and can thus be identified.
The current algorithm is based on the tracks reconstructed in the emulsions at the neutrino interaction. 
Each emulsion track is projected on the last three muon detector planes, comprising both horizontal and vertical bars providing $X$-$Y$ coordinates. The corresponding $X$-$Y$ bars hit by the projected track on each of the three planes are identified. The primary track is considered found on a given plane if there is a corresponding hit in the bars corresponding to the projected track location or in the adjacent bars. Isolation of the track is determined by requiring no hits in the bars adjacent to the search window.

A muon candidate is selected according to the following criteria:

\begin{itemize}
    \item A corresponding track is found on all the three most downstream muon detector planes.
    \item The track is isolated at least in two of the three planes.
\end{itemize}

The above selection results in a muon identification efficiency of about 73$\%$ for charged current $\nu_\mu$ interactions. Accounting also for the geometrical acceptance described above, the overall muon identification efficiency is $\sim$ 65$\%$ for $\nu_{\mu}$ CC.
The probability of misidentifying a primary hadron track as a muon was studied using neutral current (NC) $\nu_\mu$ interactions and results to be less than 0.31$\%$. 
Having on average 5 charged hadron tracks per event, this implies a purity in muon identification of almost 99$\%$.
Table~\ref{tab:muIDresults} reports the event classification for CC and NC muon neutrino interactions. 0$\mu$ events are those without any muon reconstructed in the final state. Events with one or more muons reconstructed are labelled as 1$\mu$ or $n\mu$ events. As it can be seen, 73\% of the CC interactions shows at least one reconstructed muon  while the probability for a NC interaction to be properly classified as 0$\mu$ is above 99\%.

\begin{table}
\centering
\begin{tabular}{c c c}
\toprule
 & $\%$ evts & $\%$ evts \\
 &  CC-DIS &  NC-DIS \\
 \midrule
  $0\mu$  & 37.8 & 99.7\\
  $1\mu$  & 71.5 & 0.3\\
  $2\mu$  & 1.3  & 0.03\\
  $3\mu$  & 0.1  & 0.01\\
  $>3\mu$ & 0.05 & 0.01\\
   \bottomrule
 \end{tabular}
 \caption{Event classification for CC and NC interactions. }
  \label{tab:muIDresults}
 \end{table}

The estimated muon identification efficiency can be improved not only with a better offline selection, but also moving to a muon system where the scintillating bars of the three most downstream planes in the muon system are replaced by tiles, as foreseen for the extended run configuration.

\paragraph{Electron identification}

Electrons produced by neutrino scattering are identified by the observation of the electromagnetic shower induced inside the interaction brick. 

Electromagnetic showers originated  by photons are separated from those initiated by electrons thanks to the micrometric accuracy of the nuclear emulsion that is capable of observing the displaced vertex associated with the photon conversion. This procedure is characterized by an efficiency larger than 95\%. 
    
The SciFi detector will complement the emulsion detector in the identification of electromagnetic showers and in the electron/$\pi^0$ separation, thus achieving an efficiency of about 99\%.

\paragraph{Tau identification}
The selection of the $\tau$ lepton candidates is based on purely topological criteria.
Once the primary neutrino interaction vertex has been defined, possible secondary vertices, sign of possible short lived particle decays, are searched for. 
This is done by a decay search procedure: tracks are defined as belonging to a secondary vertex if the impact parameter of the daughter track with respect to the primary vertex is larger than 10 $\mu$m.
The fraction of events in which the tau lepton decays before the last five emulsion plates is: $\epsilon_{\tau length} = 74.8 \pm 0.7\%$.

The decay search efficiency $\epsilon_{ds}$ ranges from 80$\%$ to 89$\%$ for the different channels as it is summarized in Table~\ref{tab:DSeff}. 
The total efficiency $\epsilon_{tot}$, also reported in Table~\ref{tab:DSeff}, is the combination of  geometrical,  location  and  decay search efficiencies: $\epsilon_{tot} = \epsilon_{geom} \cdot \epsilon_{loc} \cdot \epsilon_{\tau length} \cdot \epsilon_{ds}$ and ranges from 48.4$\%$ in the $\tau \to h$ decay channel up to $54\%$ in the $\tau \to 3h$ decay channel.

\begin{table}
\centering
\begin{tabular}{c  c c  c}
\toprule
decay channel & $\epsilon_{ds}$ ($\%$) & $\epsilon_{tot}$ ($\%$)\\
 \midrule
  $\tau \rightarrow \mu$ &  82.5 $\pm $ 1.6 & 49.6 $\pm$ 1.8\\
   $\tau \rightarrow e$ & 80.8 $\pm$ 1.7 & 48.4 $\pm$ 1.8\\
  $\tau \rightarrow h$ & 80.3 $\pm$ 1.0 & 48.4 $\pm$ 1.1\\
  $\tau \rightarrow 3h$ & 89.4 $\pm $ 1.5 & 54.0 $\pm$ 1.9\\
   \bottomrule
 \end{tabular}
 \caption{Decay search and overall efficiencies for the different $\tau$ decay channels}
  \label{tab:DSeff}
 \end{table}

\subsubsection{Background in the tau neutrino search}

Tau neutrino interactions are identified with the observation of two-vertices events.
Hadron re-interactions and charmed particles decays may mimic this topology. 
The hadronic re-interaction background can however be strongly reduced requiring a single or a three-prong topology and the absence of nuclear fragments  up  to  tan$\theta$ = 3.

\begin{figure}
 \centering
  \subfloat[]{\includegraphics[height = 0.35\textwidth]{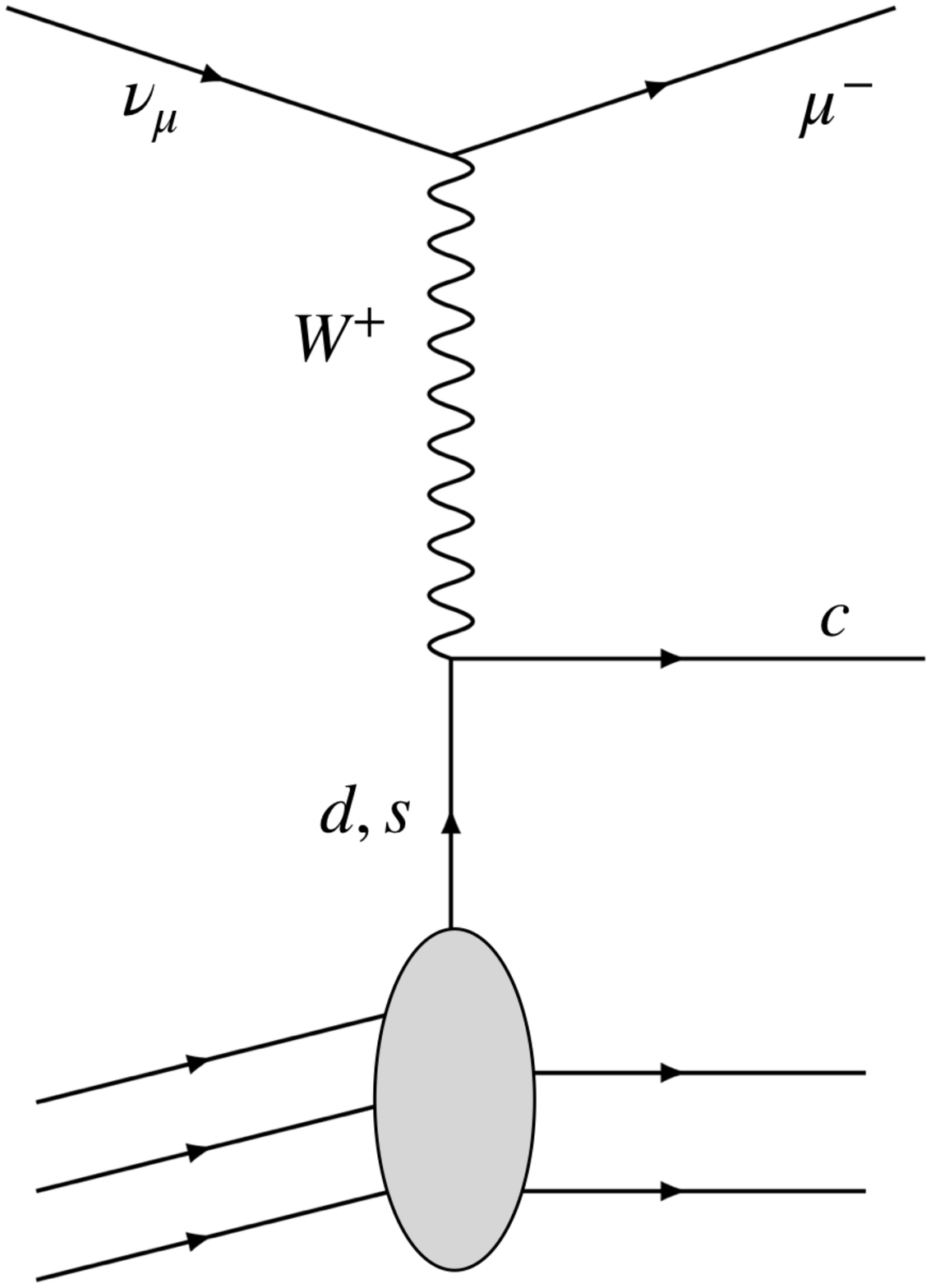}\label{sf:charmF}} \quad\quad
\subfloat[]{\includegraphics[height = 0.35\textwidth]{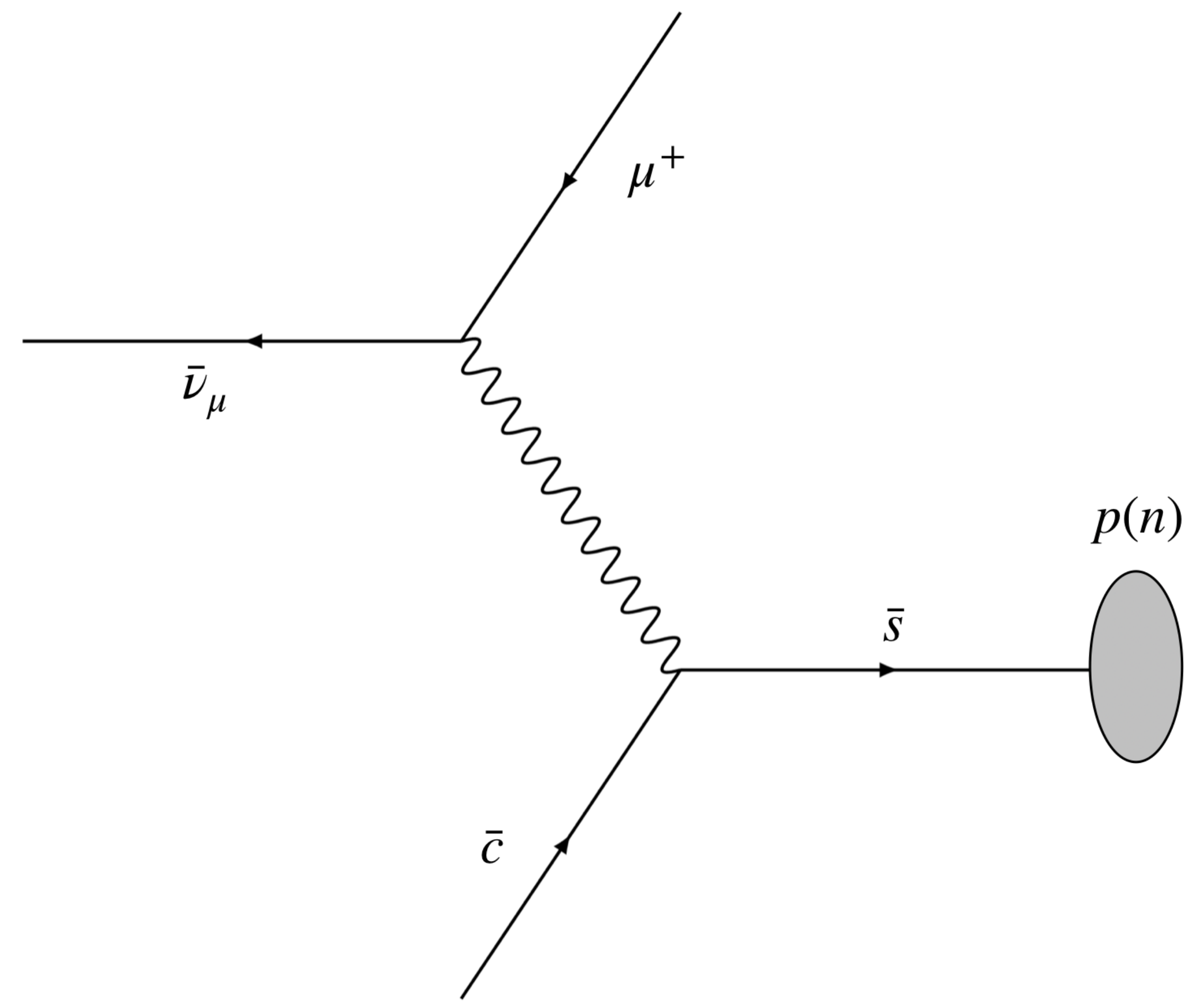}\label{sf:antichF}}
 \caption{Charm production in neutrino (a) and anti-neutrino (b) charged-current interactions.}
 \label{fig:charmFeynman}
 \end{figure}

Charmed hadrons are produced at a level of a few percent in high energy neutrino and anti-neutrino charged-current interactions~\cite{Lellis:2004yn} via the process reported in Fig.~\ref{fig:charmFeynman}.
When the  lepton produced in the  neutrino charged current interaction is not identified, the $\nu_\mu (\bar{\nu}_\mu)$ and $\nu_e (\bar{\nu}_e)$ interaction with subsequent charmed hadron production are the main background to $\nu_\tau$  searches. The  expected  charm  yields and the corresponding fractions w.r.t.~the  neutrino  charged  current  interactions for first are reported in table~\ref{tab:NuCCcharm}.

\begin{table}
\centering
\begin{tabular}{c  c c  c c}
\toprule
Neutrino &    $\langle\text{E}\rangle$  & Charm CC-DIS &  $N_{charm}/N_{CC}$\\
 flavour &   GeV    & Interactions & ($\%$)\\
\midrule
  $\nu_\mu$ & 522 &  6 &  9.7 \\
   $\nu_e$     & 751 & 2 & 9.8 \\
  $\bar{\nu}_\mu$ & 593 & 3 &  9.6 \\
   $\bar{\nu}_e$  & 775  & 1 & 10.8\\
\midrule
 TOT & & 12 & 9.8\\
\bottomrule
 \end{tabular}
 \caption{(Left column) Expected number of CC-DIS interactions with subsequent charm production for the different neutrino flavours in the assumption that 25 fb$^{-1}$ are collected in the initial detector configuration. (Right column) Fraction with respect to the total number of CC-DIS interactions.}
  \label{tab:NuCCcharm}
 \end{table}
 
Given the muon identification efficiency, the expected number of charmed hadron  events induced from $\nu_\mu$ ($\bar{\nu}_\mu$) is 3. 
This background can be further reduced with dedicated kinematical analysis~\cite{Agafonova:2018auq}.

\subsubsection{Reconstruction of neutrino energy}

We exploit the information of the electronic detectors in both the target region and in the muon identification system. Indeed, the whole detector can be  considered as a non-homogeneous calorimeter.
The energy of the hadronic jet of the neutrino interaction can be reconstructed as:

\begin{equation}
    E^{rec}_{had} = A + B\times N\text{hit}_{\text{SciFi}}+ C\times N\text{hit}_{\text{MuFilter}}
\end{equation}

\begin{figure}[h]
\centering
\includegraphics[width = 0.5\textwidth]{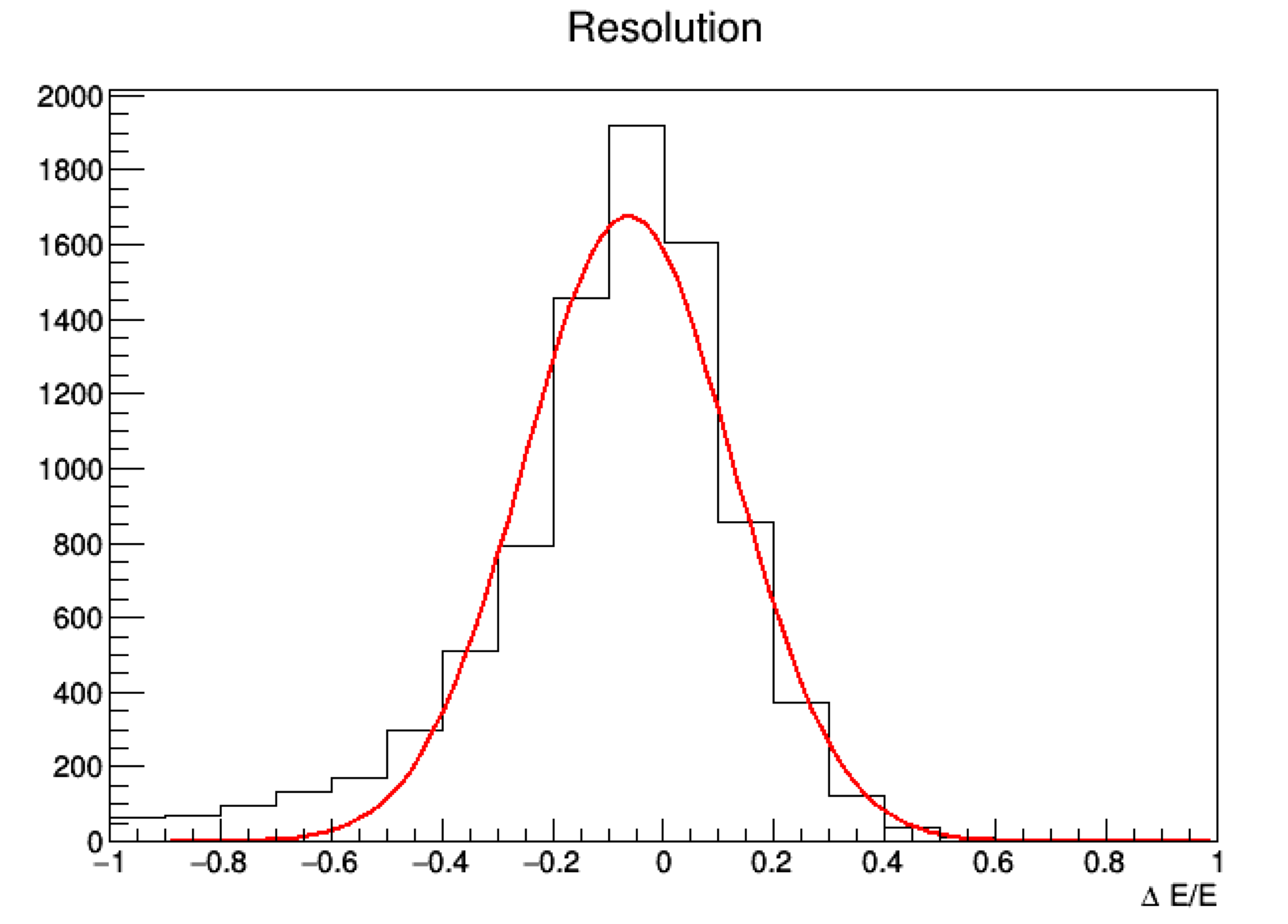} 
\caption{$\Delta E/E$ distribution of the reconstructed hadronic energy.}
\label{fig:HadERes}
\end{figure}

The values of the parameters A, B and C are obtained by a gradient descent minimisation algorithm applied to simulated events, with the following cost function:

\begin{equation}
    J(A,B,C) = J(\underline{\theta})= \frac{1}{2m} \sum_{i=1}^m {(E^{rec \, (i)}_{had}-E^{true \, (i)}_{had})^2}
\end{equation}
where $m$ is the number of events. The best-fit parameter values are: A = 1, B = 0.37, C = 0.44. 
The resolution on the reconstructed hadronic energy is reported in Fig.~\ref{fig:HadERes}.
The fractional resolution is  $\sigma(E_{had}) = (18.8 \pm 0.2) $\%. 

For $\nu_\mu$ ($\bar{\nu}_\mu$) charged current interactions, the momentum of the outgoing muon can be estimated by balancing the transverse momentum of the hadronic system. Dedicated algorithms based on multivariate techniques will be used to extract the neutrino energy. 

\subsection{Light Dark Matter search}
\label{subsec:LDM}
The increasing interest in the understanding of the nature of Dark Matter has been lately accompanied by a corresponding effort of the scientific community, both in direct and indirect searches as described in Sec.~\ref{sec:phmotivation}. Complementary approaches such as the direct observation of Dark Matter scattering off electrons have been conceived, as proposed by the SHiP experiment at the CERN SPS~\cite{Alekhin:2015byh, Ahdida:2654870}.

The SND@LHC experiment here presented is capable of performing model-independent searches for FIPs, thus exploring a large variety of the Beyond Standard Model (BSM) scenarios describing Hidden Portals. Among the most compelling ones, the paradigm of a Vector Portal which, in a minimal SM extension, reports the production of a vector mediator \textit{Dark Photon} $\mathcal{A^\prime}$ (DP):
\begin{equation}
    \mathcal{L}_{\mathcal{A}^{\prime}} =- \frac{1}{4} F'_{\mu \nu}F^{\prime \mu \nu} +\frac{m^2_{\mathcal{A}\prime} }{2}A^{\prime \mu} A^{\prime}_{ \mu}-\frac{1}{2} \epsilon F^{\prime}_{\mu \nu} F^{\mu \nu}\,,
\end{equation}
which is kinetically mixed with the photon field $F^{\mu\nu}$ via the coupling $\epsilon$.

For this study we consider prompt decays of DP into a pair of Light Dark Matter (LDM) candidates $\chi$ (i.e. $m_{\chi}\sim O(1\,$GeV/c$^2$) either fermionic or scalar, charged under a new $U^{\prime}$(1) symmetry:
\begin{equation}
    \mathcal{L}_{\mathcal{\chi}} = g_{D}\,A^{\prime\mu}\times \begin{cases} 
   \bar{\psi}_{\chi}\gamma_{\mu}\psi_{\chi} \\
   i[(\partial_{\mu}\phi_{\chi}^{\dagger})\phi_{\chi}-\phi^{\dagger}\partial_{\mu}\phi_{\chi}]
   \end{cases}
\end{equation}
being $\psi_{\chi}$ and $\phi_{\chi}$ the fermionic and scalar candidates, respectively; the parameter $g_{D}$ denotes the gauge coupling of the introduced dark sector $U^{\prime}$.\\
In the regime $m_{A^{\prime}}>2\,m_{\chi}$ (invisibly decaying DP) and $\alpha_{D}=g_{D}^{2}/4\,\pi\, \gg\epsilon\,e$, it is reasonable to assume a short-lived DP with $\mathrm{BF}(A^{\prime}\to\chi\chi^{\dagger})\approx 1$.

At the SND@LHC experiment, a multitude of processes give rise to the production of DPs and, consequently, LDM particles. In the considered mass range $m_{\mathcal{A}^\prime}\lesssim\,$1 GeV, two mechanisms are leading:
\begin{itemize}
\item \textit{Meson decays}: DPs abundantly originate from radiative decays of light mesons in
\[
\pi^{0},\,\eta,\,\eta^\prime\to\gamma\mathcal{A}^{\prime},
\]
\[
\omega\to\pi^0\mathcal{A}^{\prime}\,.
\]
\item \textit{Bremsstrahlung of protons}: interacting primary protons at the LHC collision point radiate DPs at a very low angular spread.
\end{itemize}

It is here noted that on-shell production of DPs from prompt-QCD interactions contributes to a negligible extent, thus ignored for this study.

The experimental signature considered in this study is the LDM elastic scattering off the detector atomic electrons:
\[
\chi\,e^{-}\to\chi\,e^{-}
\]
where the recoil electron-induced shower is reconstructed in the target region with high position resolution and angular precision, thanks to its high-granularity. LDM elastic scattering off nuclei and inelastic scattering represent an appealing perspective and are left as subject for future investigation.

Possible backgrounds to this search consist of neutrino interactions where exclusively one charged track, either an electron or positron, is reconstructed at the primary vertex.
Different contribution channels have been investigated, namely: neutrino-electron CC/Neutral Current (NC) scattering $\nu_{x}\,e^{-}\to\nu_{x}\,e^{-}$, exhibiting the same topology of the signal; $\nu_e(\bar{\nu}_{e})$ CC Deep Inelastic Scattering (DIS), CC Resonant Scattering (CCRES) and CC Quasi-Elastic Scattering (CCQE) with soft undetectable tracks at the neutrino vertex.

Full Monte Carlo simulations performed by means of the software packages described in Sec.~\ref{subsec:neutrino} for the incoming neutrino flux and by  Genie~\cite{cite:GENIE} for the neutrino scattering, have shown that the background comes exclusively from CCDIS interactions, given the negligible yield of other sources. Once visibility criteria (Tab.~\ref{tab:VisCut}) are applied to the tracks produced in CCDIS interactions, the corresponding yield becomes negligible. Zero background is therefore assumed in the following analysis.

\begin{figure}[htb]
\centering
    \includegraphics[width=.6\textwidth]{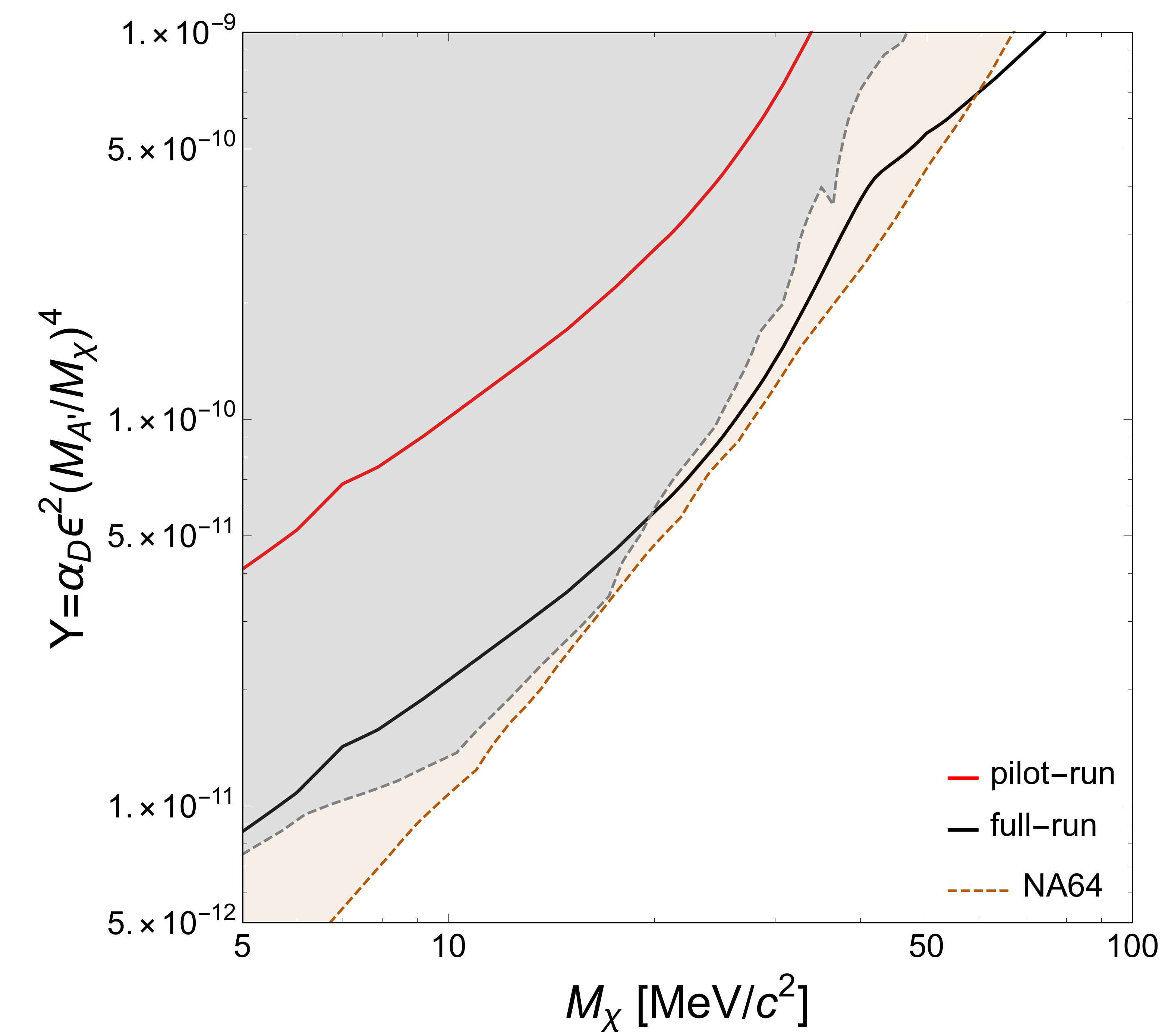}
    \caption{SND@LHC $90\%$ C.L. exclusion limits in the 0-background scenario for a LDM candidate $\chi$ originated from the prompt decay of a DP $\mathcal{A^\prime}$, assuming as benchmark parameters $m_{\mathcal{A}^\prime}=3\,m_{\chi}$ and $\alpha_{D}=0.1$.}\label{fig:sens_lowmass}
\end{figure}

A full MC simulation has been performed also for the signal. LDM candidates have been produced by means of {\sc Pythia8}~\cite{Pythia8,Sjostrand:2014zea} and MadDump~\cite{maddump} MC generators, the latter tuned to describe the current detector configuration and model physical parameters: a benchmark scenario is assumed, in which $m_{\mathcal{A}^\prime}=3\,m_{\chi}$ and $\alpha_{D}=0.1$. We report in Fig.~\ref{fig:sens_lowmass} the projected $90\%\,$C.L. exclusion limits in the plane (M$_{\chi}$, Y$=\epsilon^{2}\alpha_{D}\bigl(M_{\chi}/M_{A^\prime}\bigr)^{4}$) for the first run and the full data-taking period ($\sim\!25$ and 150 fb$^{-1}$, respectively), compared to existing constraints.

\begin{figure}[htb]
\centering
    \includegraphics[width=.75\textwidth]{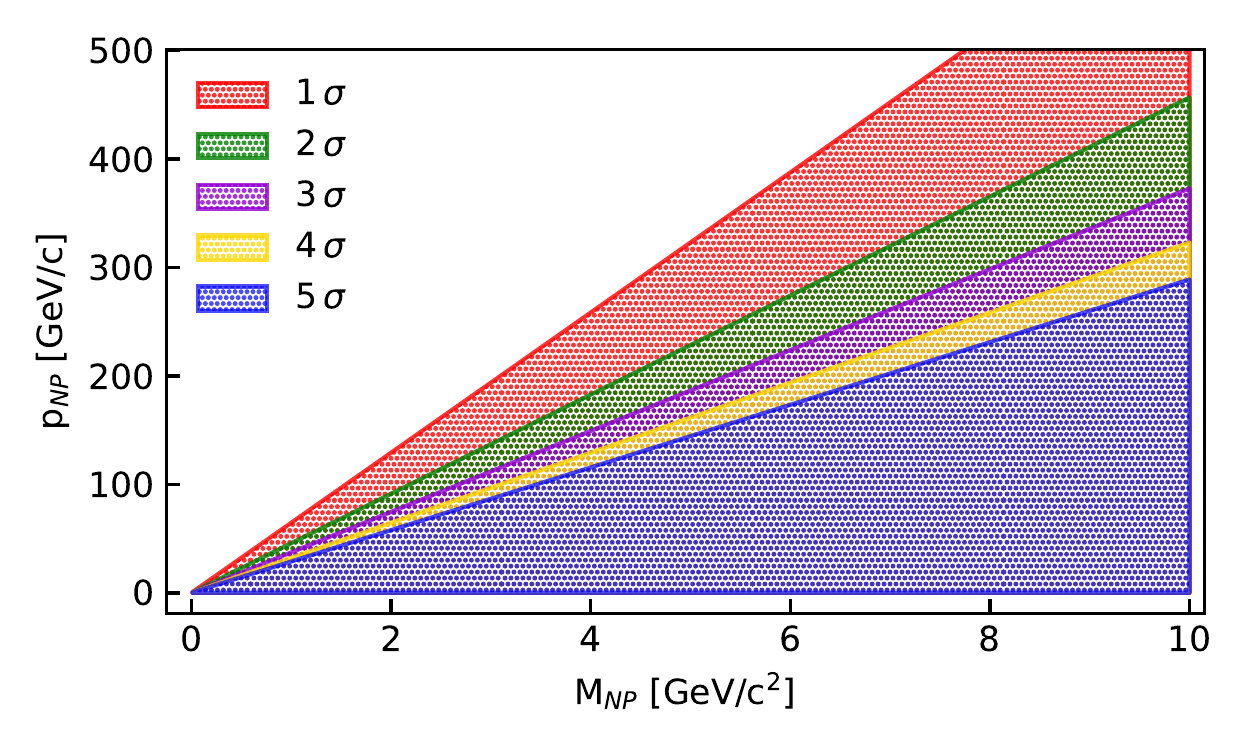}
    \caption{Sensitivity curves for a TOF measurement, illustrated from 1 up to $5\sigma$ in the plane ($M_{\mathrm{NP}}$, $p_{\mathrm{NP}}$) of the New Particle candidates to be detected in the SND@LHC detector, produced assuming a time resolution of $\sim\!200\,$ps. }\label{fig:sensTOF}
\end{figure}

A complementary approach consists of using the Time Of Flight (TOF) measurement, once the time information  is matched to the ECC. With a time resolution of $\sim\!200\,$ps, 
it will be possible to disentangle any FIP scattering process w.r.t.~neutrino ones, 
with a significance which depends on the particle mass. The exploitable phase space of this proposal is shown in terms of sensitivity curves from 1 to $5\sigma$ in the plane $(M_{\mathrm{NP}},\,p_{\mathrm{NP}})$ in Fig.~\ref{fig:sensTOF}, where $M_{\mathrm{NP}}$ and $p_{\mathrm{NP}}$ denote the mass and momentum, respectively, of the new physics candidate to be distinguished from ordinary neutrinos. The $1\sigma$ contour in Fig.~\ref{fig:sensTOF} corresponds to a $\gamma$~factor of about 50, while the $5\sigma$ one to a $\gamma$~factor of about 25.

The physics case can be extended by adding the scattering with nuclei and other possible models.
\section{Installation and operation plan}

\begin{figure}[hbtp]
  \centering
  \includegraphics[width=0.9\linewidth]{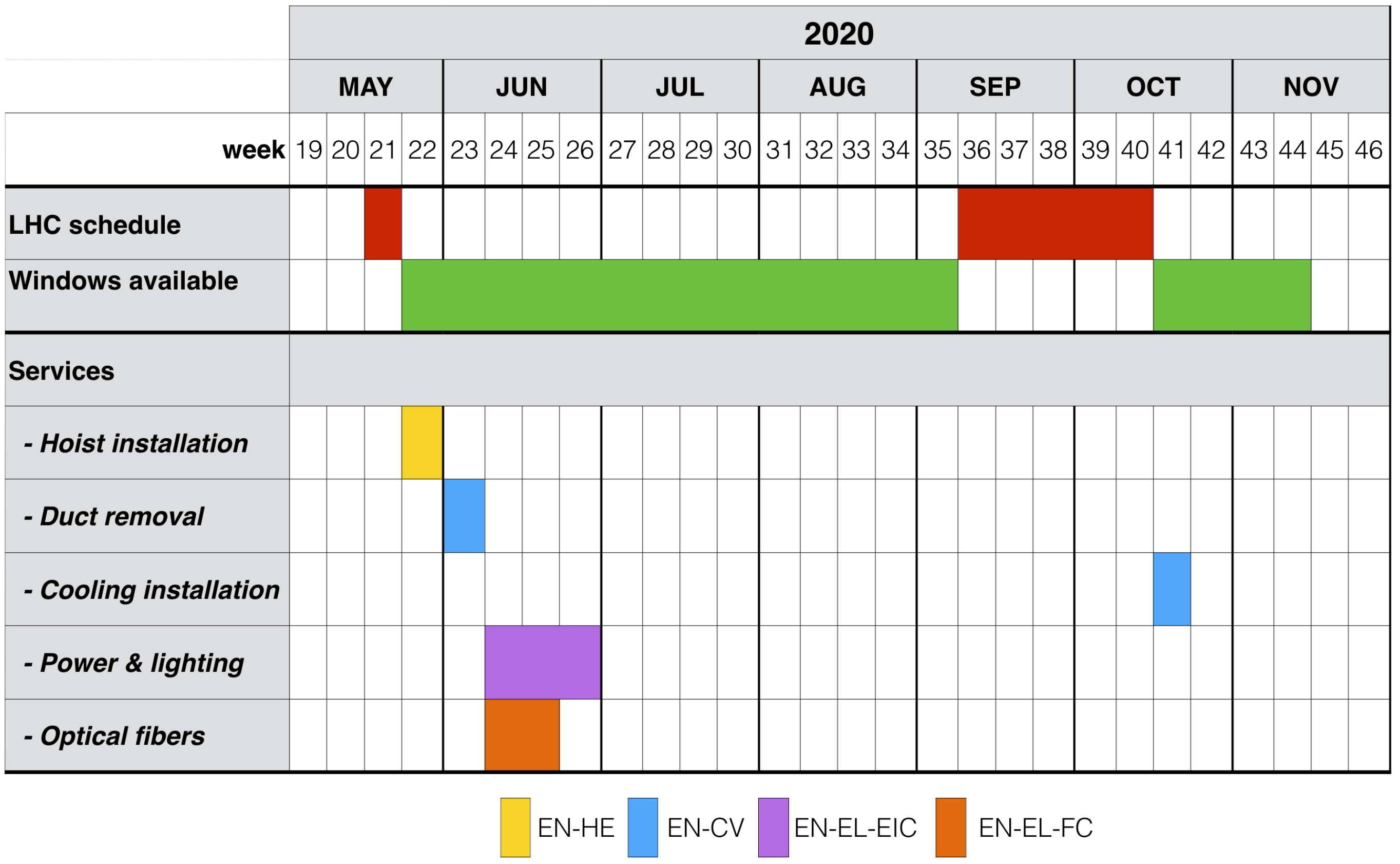}
  \caption{Preliminary plan for the installation of the services in TI18.  }
  \label{fig:planning}
 \end{figure}

A preliminary schedule for the installation of all the needed services, cross-checked with EN-ACE-OSS and based on the LHC Long Shutdown 2 Planning~\cite{LHCschedule},  is reported in Fig.~\ref{fig:planning}. The  duration of each task for the installation of services in TI18 has been estimated and the corresponding CERN groups identified. 
The timeline reflects preliminary discussions with the different groups. As an example, EN-CV has already pointed out that the chillers need to be previously commissioned on the surface before installing them in the LHC tunnel. Six months are needed from the design phase to their final installation in TI18. Until week number 44, the responsible of the LHC machine coordination is EN-ACE-OSS and the available windows for installation are well identified (highlighted in green in the table). 
Access will also be possible until the end of LS2 to finalise installation and commissioning but it will require careful coordination with BE-OP-LHC as the schedule for the LHC machine tests is not yet fixed. 

The installation of services can be completed by October 2020. At the same time, the SciFi stations  and the Veto detector will be ready. Emulsion bricks will be installed at the very last moment, before the LHC machine starts. Additional scintillating bar planes for the muon system will be constructed in the meanwhile. We expect that four planes could be ready to be installed within mid 2021. This number would be sufficient to get either a calorimetric measurement of the energy or the muon identification, depending on their location. The optimal configuration  as a function of the actual number of available scintillating bar planes will be worked out.  The installation of the remaining planes of scintillator bars will be completed by 2021.

\section{Summary}\label{sec:summary}
With this paper, the SHiP Collaboration expresses its interest in the construction and operation of a detector in the TI18 tunnel at the LHC that, for the first time, will measure the process $p p \rightarrow \nu X$ in the pseudorapidity region $7.2 < \eta < 8.7$, where neutrinos are mostly produced from charm decays, and search for FIPs in an unexplored domain. The proposed detector is hybrid, combining the nuclear emulsion technology and electronic detectors. This combination results in a very compact and high-performance neutrino detector capable of identifying neutrino interactions of the three flavours and perform searches for neutral massive particles via their scattering on the detector material. The detector is a prototype of the Scattering and Neutrino detector of the SHiP experiment. These measurements will also provide important input to the optimization of the SHiP detector in the years to come. 

In 2018 the SHiP Collaboration performed an exposure of an emulsion-based detector to $1.5 \times 10^6$ protons at the H4 beamline on the SPS. This exposure aimed at measuring charm production induced by 400~GeV proton interactions in a thick target where the cascade production of charmed hadrons is relevant. This measurement has triggered the development of new reconstruction software in the challenging conditions where proton interactions have to be reconstructed in an environment with a large flux of other particles, with an occupancy in the emulsion up to $5 \times 10^4$ particles/cm$^2$. The results show a good agreement between data and the Monte Carlo simulation both in normalization and in shape, thus demonstrating the feasibility of  reconstructing neutrino interactions  in the context of the measurement we propose here. 

An integration study conducted with all the relevant engineering departments at CERN has not shown any showstopper neither in setting up the relevant services, nor in the detector installation and operation. 

With this expression of interest, the Collaboration is seeking the approval for a  run in 2021-2022 where first data will be collected.  In perspective, the Collaboration could take data throughout the Run 3 of the LHC. 

\section*{Acknowledgments}\label{sec:Acknowledgments}
We thank our colleagues Yannis Karyotakis and Jocelyn Monroe for 
 fruitful discussions during the preparation of this manuscript.

\bibliographystyle{JHEP} 
\bibliography{references}

\end{document}